\documentclass{aa}

\input psfig.tex

\usepackage{graphicx}
\usepackage{natbib}
\usepackage{array}
\bibpunct{(}{)}{;}{a}{}{,}

\usepackage{rotating}  

        \newcommand{\Hbeta}{\mbox{${\rm H}\beta$}}
        \newcommand{\MgFe}{\mbox{${\rm [MgFe]} $}}
        
        \newcommand{\MFe}{\mbox{${\rm \langle Fe \rangle} $}}
        \newcommand{\mgii}{\mbox{${\rm Mg2}$}}
        
        \newcommand{\mgb}{\mbox{${\rm Mgb}$}}
        \newcommand{\nad}{\mbox{${\rm NaD}$}}
        
        \newcommand{\hga}{\mbox{${\rm H}\gamma {\rm A}$}}
        \newcommand{\hda}{\mbox{${\rm H}\delta {\rm A}$}}
        \newcommand{\hgf}{\mbox{${\rm H}\gamma {\rm F}$}}
        \newcommand{\hdf}{\mbox{${\rm H}\delta {\rm F}$}}

    \newcommand{\FeH}{\mbox{[Fe/H]}}

    \newcommand{\xsfe}{\mbox{[$X_{{\rm el}}$/Fe]}}
    \newcommand{\asfe}{\mbox{[$\alpha$/Fe]}}
    \newcommand{\gaa}{\mbox{$\Gamma_{\alpha}$}}
        \newcommand{\alfa}{\mbox{$\alpha$-elements}}
        \newcommand{\alfe}{\mbox{$\alpha$-enhanced}}

    \newcommand{\Teff}{\mbox{$T_{{\rm eff}}$}}  
    \newcommand{\logT}{\mbox{$\log T_{{\rm eff}}$}}
    \newcommand{\logG}{\mbox{$\log g$}}

        \def\smallskip{\vskip 4pt}
        
\begin{document}

\title{New response functions for absorption-line indices from high-resolution spectra}
\subtitle{}
     \author{Rosaria Tantalo\inst{},
           Cesare Chiosi \inst{},  \&
           Lorenzo Piovan \inst{}
            }
     \offprints{R. Tantalo }
     \institute{ Department of Astronomy, University of Padova,
                Vicolo dell'Osservatorio 2, 35122 Padova, Italy \\
                \\
\email{tantalo@pd.astro.it; chiosi@pd.astro.it;
        piovan@pd.astro.it
      }
      }
     \date{Received: June, 2005; Revised: November 2005, June 2006; Accepted: }

\abstract{ Basing on the huge library of 1-\AA\ resolution spectra
calculated by Munari et al. over a large range of \logT, \logG, [Fe/H]
and both for solar and \alfe\ abundance ratios [$\alpha$/Fe], we
present theoretical absorption-line indices on the Lick system.  First
we derive the so-called response functions ($\mathcal{RF}$s) of
Tripicco \& Bell for a wide range of \logT, \logG, [Fe/H] and
\asfe=+0.4\,dex. The $\mathcal{RF}$s are commonly used to correct
indices with solar \asfe\ ratios to indices with \asfe$>$0. Not only
the $\mathcal{RF}$s vary with the type of star but also with the
metallicity. Secondly, with the aid of this and the fitting functions
($\mathcal{FF}$s) of Worthey et al., we derive the indices for single
stellar populations and compare them with those obtained by previous
authors, e.g. Tantalo \& Chiosi. The new $\mathcal{RF}$s not only
supersede the old ones by Tripicco \& Bell, but also show that \Hbeta\
increases with the degree of enhancement in agreement with the results
by Tantalo \& Chiosi. The new indices for single stellar populations
are used to derive with aid of the recursive Minimum Distance method
the age, metallicity and degree of enhancement of a sample of Galactic
Globular Clusters for which these key parameters have been
independently derived from the Colour-Magnitude Diagram and/or
spectroscopic studies. The agreement is remarkably
good.  }

\titlerunning{Absorption-line indices from high-resolution spectra}
\authorrunning{R. Tantalo et al.}
\maketitle

\section{Introduction}\label{intro}

The Lick system of absorption-line indices developed over the years by
\citet{Burstein84}, \citet{Faber85}, \citet{Worthey92}, \citet{Worthey92b},
\citet{Worthey94} and \citet{Worthey94a} was designed to infer the age and
the metallicity of stellar systems, early-type galaxies in particular.

The Lick indices seem to have the potential of partially resolving the
age-metallicity degeneracy that is long known to affect the spectral
energy distribution of stellar populations
\citep{Renbuz86}\footnote{An old metal-poor stellar population may
happen to have the same spectral energy distribution of a young
metal-rich one.}. Thanks to it, an extensive use of the Lick system of
indices has been made by many authors
\citep{Bressan96,Tantalo98,Tantalo98a,Trager20a,Trager20b,Kuntschner98,
Kuntschner98a,Jorgensen99,Kuntschner00,Poggianti01,Kuntschner01b,
Vazdekis01,Davies2001,Maraston03,Thomas03,Thomas03a,ThoMara03,Tantalo04a}.
The main result of all those studies is that even in early-type
galaxies some recent episodes of star formation ought to occur in
order to explain the large scatter shown by the observational data for
indices like \Hbeta\ commonly thought to be sensitive to the turn-off
stars and consequently to the age.

The problem is, however, further complicated by a third parameter,
i.e. the abundance ratios \asfe\ (where $\alpha$ stands for all
chemical elements produced by $\alpha$-captures on lighter nuclei).
Absorption-line indices like \mgii\ and \MFe\ measured in the central
regions of galaxies are known to vary passing from one galaxy to
another \citep{Gonzalez93,Trager20a,Trager20b}. Looking at the
correlation between \mgii\ and \MFe\ (or similar indices) for the
galaxies in the above quoted samples, \mgii\ increases faster than
\MFe, which is interpreted as due to enhancement of \alfa\ in some
galaxies. In addition to this, since the classical paper by
\citet{Burstein88}, the index \mgii\ is known to increase with the
velocity dispersion (and hence mass and luminosity) of the galaxy.
Standing on this body of data the conviction arose that the degree of
enhancement in \alfa\ ought to increase passing from dwarf to massive
early-type galaxies \citep{Faber92,Worthey94,Mattfra94,Matteucci97,
Matteucci98}. These findings strongly bear on the theory of galaxy
formation as super-solar \asfe\ ratios require rather short
star-formation time-scales (see for instance the discussion of this
topic by \citet{Chiocar02}) and the correlation with the velocity
dispersion requires that the star formation time-scale should get
longer at decreasing galaxy mass, in contrast with the standard
supernova driven galactic wind model by \citet{Larson74}, and instead
supported by the N-Body-Tree-SPH models of early-type galaxies by
\citet{Chiocar02}.

In presence of \alfe\ chemical compositions, ages and metallicities of
early-type galaxies should be derived from indices in which
$\alpha$-enhancement is included. As long ago noticed by
\citet{Worthey92b} and \citet{Weiss95}, indices for \alfe\ chemical
mixtures of given total metallicity are expected to differ from those
of the standard case. On one hand, this spurred new generations of
stellar models, isochrones, SSPs with $\alpha$-enhanced mixtures
\citep{Salasnich20} and, on the other hand, led to many attempts of
increasing complexity to simultaneously derive, from fitting the
observational indices to their theoretical counterparts, the age,
metallicity and degree of enhancement \citep{Tantalo98a,Trager20a,
Trager20b,Maraston03,Thomas03,Thomas03a,ThoMara03,Tantalo04a}. The
formal solution for ages, metallicities and \asfe\ ratios based on
large samples of galaxies, e.g. the \citet{Trager98} list, once more
yields a large range of ages, metallicities and abundance ratios, as
amply discussed by \citet[][hereafter TC04]{Tantalo04a}.

Although the picture emerging from the above studies is a convincing
one, there are still several points of weakness intrinsic even to the
state-of-the-art theoretical indices that force us to reconsider the
whole problem. Let us examine in some detail (i) the foundations of
the Lick indices; (ii) the various steps that are required to derive
theoretical indices and their dependence on age, metallicity and
degree of enhancement; (iii) the current method to estimate these
parameters from the indices; (iv) and finally shortly comment on a few
points of controversy among different groups.

\begin{itemize}
\item The information contained in the stellar spectra concerning the
effective temperature (\logT), gravity (\logG) and chemical
composition (usually the parameter \FeH) has been coded in a system of
indices measuring the strength of atomic and molecular lines (see
Section~\ref{indices}). The Lick indices are defined and measured on a
finite sample of stellar spectra with fixed mean resolution of 8.4\AA\
(which in most cases differs from the one of theoretical spectra used
in population synthesis).

\item In order to make possible the general use of the Lick indices,
\citet{Worthey94} introduced the concept of the so-called fitting functions
($\mathcal{FF}$s). The indices for a large sample of stars with known
atmospheric parameters (\logT, \logG\ and \FeH) are measured and then
expressed as empirical polynomial fits as functions of these
parameters. The major drawback with the $\mathcal{FF}$s is that they
encompass a limited range of values in particular as far as \FeH\ is
concerned. The sample of stars indeed was collected from the solar
vicinity even if attempts to extend it to lower metallicities have
been made (\citet{Idiart95}, \citet{Cenarro01,Cenarro02}).
Metallicities much higher than those in the solar vicinity are simply
not included for obvious reasons. A new library is also available with
unprecedented coverage of atmospheric parameters by
\citet{Sanchez04a} and Sanchez-Bl\'azquez et al. (in preparation).

\item Despite it was long known that magnesium is perhaps enhanced with
respect to iron ([Mg/Fe]$>$0) in giant elliptical galaxies (see e.g.
\citet{Worthey92b} and above), the collection of stellar spectra and
$\mathcal{FF}$s in turn did not allow for non-standard abundance
ratios in the chemical composition. Only occasionally and for a few
indices, e.g. \mgii\ and NaD, $\mathcal{FF}$s including the effect of
non-standard abundance ratios have been proposed, e.g. \citet{Borges95}.

\item A milestone along the road was put by \citet[][]{Tripicco95} who have
modeled synthetic spectra of high-resolution (0.1\AA), degraded them
to the mean 8.4\AA\ resolution of the Lick system, and derived from
them the indices for three prototype stars, namely a cool-dwarf, a
turn-off and a cool-giant along a 5\,Gyr isochrone matching the CMD of
M67 (in other words for three different combinations of \logT\ and
\logG). Even more important here, they studied the response of the
indices produced by changes in the abundance ratios of individual
elements. In other words, thanks to their $\mathcal{RF}$s it was
possible to evaluate the effect of abundance ratios different from
solar.

\item With the aid of the \citet{Tripicco95} $\mathcal{RF}$s and suitable
algorithms, the indices with solar abundance ratios have been
transformed into those for non solar abundance ratios. The algorithms
in use are not unique thus leading to uncertain results (see TC04 for
more details).

\item Passing now from individual stars to star clusters, reduced here
to single stellar populations (SSPs), and galaxies (manifolds of
SSPs), the derivation of theoretical indices (and spectra, magnitudes
and broad-band colours) is even more complicate because other
ingredients intervene: (i) the construction of realistic isochrones
for SSPs including all evolutionary phases (even the unusual ones that
are known to appear at very high metallicities, see \citet{Bressan94}
and TC04); (ii) the initial mass function; (iii) and finally, in the
case of galaxies, the past history of star formation and chemical
enrichment weighing the contribution from stellar populations of
different age and chemical compositions \citep[see e.g.][for all
details]{Bressan94}. It is worth commenting here that galaxy indices
are almost always compared to SSP indices thus neglecting the mix of
stellar populations and missing important contributions from some
peculiar components. For instance an old SSP of very high metallicity
and/or a very old SSP of extremely low metal content, would possess
strong \Hbeta\ thus mimicking a young SSP of normal metallicity (see
TC04 for more details). Integrated indices for model galaxies have
been occasionally calculated and used \citep{Tantalo98}, but never
systematically applied to this kind of analysis. This is a point that
should be carefully investigated and kept in mind when comparing data
with theory.
\end{itemize}

If for solar abundance ratios the dependence of the indices on age and
metallicity is currently on a rather solid ground but for the effect
of some unusual phases of stellar evolution, the same does not happen
for non-solar abundance ratios because it is still highly
controversial how some indices (\Hbeta, in particular) at given age
and total metallicity would respond to changes in the abundance
ratios. According to TC04, SSPs of the same age and metallicity but
different degrees of enhancement in \alfa\ may have significantly
different \Hbeta: in brief keeping age and metallicity constant,
\Hbeta\ should increase at increasing degree of enhancement (see the
discussion in Section~\ref{ind_high_res} for more details). Basing on
this, \citet{Tantalo04b} suggested an alternative, complementary
interpretation of the large scatter in \Hbeta\ shown by early-type
galaxies of the local universe in two-indices diagnostic planes like
\Hbeta\ versus [MgFe]. Quickly summarizing their study: (i) most probably
the majority of galaxies (those with \Hbeta$\leq$2) are very old
objects of the same age (say approximately 13\,Gyr) but with a
different degree of enhancement. (ii) Only for galaxies with
\Hbeta$>$2, the presence of secondary star forming activity ought to
be invoked. High-resolution synthetic spectra for stars with good
coverage of gravity, effective temperature, chemical composition and
degree of enhancement in $\alpha$-element can greatly alleviate the
above difficulties and shed light on these important issues. In
principle, indices can be straightforwardly calculated from the
spectra with no need of $\mathcal{FF}$s, $\mathcal{RF}$s and suitable
algorithms to take $\alpha$-element enhancement into account. In this
paper, we limit ourselves to present new $\mathcal{RF}$s and a simple
correction algorithm. Full exploitation of the possibilities offered
by the high resolution spectra is left to a forthcoming study.

The plan of the paper is as follows. Section~\ref{indices} deals with
the theory of absorption-line indices; Section~\ref{def_ind}
summarizes their definition for a single star; Section~\ref{enha}
presents two different methods for enhancing the abundance of \alfa\
with respect to that of Fe, i.e. at constant or increasing the total
metallicity, and examines in some detail the implications.  In this
paper we adopt the procedure at increasing metallicity for the sake of
internal consistency with the library of stellar spectra in
usage. Section~\ref{worthey_ff} recalls the dependence of the
$\mathcal{FF}$s of the Lick system on the main stellar parameters,
comments on the fact that they are based on solar-scaled chemical
compositions and shortly describes the $\mathcal{FF}$s by
\citet{Borges95} in which for some indices the effect of enhancing
$\alpha$-elements is taken into account; finally Section~\ref{tribell}
shortly reviews the method of the $\mathcal{RF}$s by
\citet{Tripicco95} and current algorithms that are adopted to transfer
solar-scaled indices into the corresponding ones with
$\alpha$-enhancement. In Section~\ref{new_calib} we present our new
$\mathcal{RF}$s. We start presenting in Section~\ref{high_res_sp} the
library of synthetic spectra in use.  The covered \logT\ and \logG\
extend nearly across the whole HR-diagram of real stars, the
metallicity goes from ${\rm [Z/Z_{\odot}]}$=--2.0 to 0.5 for spectra
with \asfe=0 and up to 0.0 for spectra with \asfe=+0.4\,dex. Secondly,
we calculate indices and $\mathcal{RF}$s. The analysis is made in two
steps: (i) adopting the definition of the indices on the Lick system
and the \citet{Tripicco95} definition of $\mathcal{RF}$s but using the
stellar spectra at 1-\AA\ resolution, we derive the indices for the
calibrating stars (Section~\ref{ind_high_res}). These indices
constitute the backbone of the new grid of \citet{Tripicco95}-like
$\mathcal{RF}$s given in Section~\ref{tb95_like}. Owing to the large
body of data on the Lick system, it is worth providing indices and
$\mathcal{RF}$s from spectra having the same resolution of the Lick
system. They are presented in Section~\ref{degrade_to_lick}. The new
$\mathcal{RF}$s (and indices) are now strictly equivalent to those of
\citet{Tripicco95}. Together with the classical $\mathcal{FF}$s they
can be used to pass from solar to $\alpha$-enhanced mixtures.  These
indices are immediately comparable to most of observational data in
literature. To this end we need to calculate indices for SSPs. This is
the subject of Section~\ref{ind_ssp}. We start shortly recalling the
index definition for SSPs and presenting the library of stellar
spectra with medium resolution specifically used to this purpose
(Section~\ref{ind_ssp_def}) and summarising the sources of stellar
models and isochrones (Section~\ref{mod_iso}). With the aid of the new
$\mathcal{RF}$s we calculate absorption-line indices for SSPs at
varying metallicity, degree of enhancement and age. The results are
presented in Section~\ref{ind_ssp_ff} and compared to those by
previous sources, e.g. TC04, in Section~\ref{compar}. The increase of
\Hbeta\ with the degree of enhancement is fully confirmed, thus
lending support to the conclusion reached by TC04. In
Section~\ref{data_globular_cluster}, the indices obtained with the new
$\mathcal{RF}$s are compared with the data for globular clusters. A
summary of the results of this study and some concluding remarks are
finally presented in Section~\ref{conclu}.

\section{Theory of absorption-line indices}\label{indices}

\subsection{Definition}\label{def_ind}

In the Lick system of absorption-line indices
\citep[see][]{Burstein84,Faber85,Worthey94} the definition of an index
with pass-band $\Delta \lambda$ is different according to whether it
is measured in equivalent width (EW) or magnitude (Mag)

\begin{eqnarray}
I_{l} &= &\Delta\lambda \left(1 - \frac{F_{l}}{F_{c}} \right) \,\,\,\,(EW)
\label{ind_def_ew}                                                            \\
I_{l} &= &-2.5 \log \left(\frac{F_{l}}{F_{c}} \right)\,\,\,(Mag)
\label{ind_def_mag}
\end{eqnarray}

\noindent 
where $F_{l}$ and $F_{c}$ are the fluxes in the line and
pseudo-continuum, respectively (e.g. see Fig.~\ref{diff_in_hb}). The
flux $F_{c}$ is calculated by interpolating to the central wavelength
of the absorption-line, the fluxes in the midpoints of the red and
blue pseudo-continua bracketing the line \citep{Worthey94}. The
pass-bands adopted for the 25 indices of the Lick system are taken
from \citet{Trager98}, see also \citet{Worthey94}. Because of the
finite sampling of the Lick library of template stellar spectra and
their different resolution with respect to theoretical spectral
libraries currently in usage, eqns.~(\ref{ind_def_ew}) and/or
(\ref{ind_def_mag}) cannot be straightforwardly applied. As already
mentioned the problem is bypassed by means of the $\mathcal{FF}$s
\citep{Worthey94}.

We also add the index ${\rm D4000}$ centered at the 4000\AA\ break. It
is defined as the ratio of the average flux $F_{\nu}$ (in ${\rm
erg~sec^{-1} cm^{-2} Hz^{-1}}$) in two bands at the long- and
short-wavelength side of the discontinuity \citep{Bruzual83}, i.e.

\begin{equation}
{\rm D4000} = \frac{\lambda_{2}^{{\rm b}}-\lambda_{1}^{{\rm b}}}{\lambda_{2}^{{\rm r}}-\lambda_{1}^{{\rm r}}}
\frac{\int_{\lambda_{1}^{{\rm r}}}^{\lambda_{2}^{{\rm r}}} F_{\nu} d\lambda}
{\int_{\lambda_{1}^{{\rm b}}}^{\lambda_{2}^{{\rm b}}} F_{\nu} d\lambda} \nonumber
\end{equation}

\noindent
The blue and red pass-bands are from 3750 to 3950\AA, and from 4050 to
4250\AA, respectively.

\subsection{$\alpha$-enhanced chemical compositions}\label{enha}

As pointed out by Bressan (2005, private communication) the current
definition of enhancement in \alfa\ of a chemical mixture is not
univocal and it may be a source of uncertainty when comparing results
from different authors and in particular when using synthetic spectra
with $\alpha$-enhanced chemical compositions. The problem can be
reduced to the statement: enhancing of \alfa\ can be made either at
constant or varying total metallicity Z.

\subsubsection{Enhancement at constant Z}\label{enha_1}

Let us take a certain mixture of elements with total metallicity Z
(sum of all elements heavier than He), denote with ${\rm N}_{j}$ the
number density of the generic element $j$ with mass abundance $X_{j}$
(${\rm N}_{j} = \rho {\rm N}_{0} X_{j}/\mathcal{A}_{j}$, where $\rho$
is the mass density, ${\rm N}_{0}$ is the Avogadro number and
$\mathcal{A}_{j}$ is the mass number) and define the quantity ${\rm
A}_{j}$

\begin{equation}
{\rm A}_{j} = \log \left(\frac{{\rm N}_{j}}{{\rm N_{H}}} \right) +12
\label{ajj}
\end{equation}

\noindent
Ignoring elements from H to He, in this mixture $\sum_{j} X_{j}={\rm
Z}$ by definition. The abundance by mass with respect to Fe is given
by

\begin{equation}
\left[\frac{X_{j}}{X_{{\rm Fe}}} \right] = \log \left( \frac{X_{j}}{X_{{\rm Fe}}} \right)
- \log \left( \frac{X_{j}}{X_{{\rm Fe}}} \right)_{\odot}
\end{equation}

\noindent
or in terms of ${\rm A}_{j}$

\begin{equation}
\left[\frac{X_{j}}{X_{{\rm Fe}}} \right] = \left( {\rm A}_{j} - {\rm A}^{\odot}_{j} \right) -
\left( {\rm A_{Fe}} - {\rm A^{\odot}_{Fe}} \right)
\end{equation}

\noindent
Keeping constant the number density of Fe, i.e. $({\rm A_{Fe}}-{\rm
A_{Fe}}^{\odot})$=0 and changing other elements (enhancement), we
obtain

\begin{equation}
{\rm A}_{j}^{\rm enh} = \left[\frac{X_{j}}{X_{{\rm Fe}}} \right] + {\rm A}_{j}^{\odot}
\end{equation}

\noindent
Inserting the adopted [$X_{j}/X_{{\rm Fe}}$] and the solar values for
${\rm A}_{j}^{\odot}$ by \citet{GrevesseSauval98}, respectively, from
equation~(\ref{ajj}) we can calculate the new abundances by mass for
any pattern of \alfe\ elements. In general, the sum of the $X_{j}$
will be different from Z. The mass abundance must therefore be
re-scaled to the true value $X_{j}'$ given by

\begin{equation}
X'_{j}  = \frac{X_{j}}{\sum X_{j}}
\end{equation}

\noindent
Finally we may define the ``total enhancement factor'' $\Gamma_{Z}$ as

\begin{equation}
\Gamma_{Z} = - \log{ \left( \frac{X'_{{\rm Fe}}}{X_{{\rm Fe}}^{\odot}} \right) }
\label{gamma}
\end{equation}

\noindent
from which we immediately obtain the relationship between $\Gamma_{Z}$
and the new value of the Fe abundance in $\alpha$-enhanced mixtures

\begin{equation}
\log \left(\frac{X'_{{\rm Fe}}}{X'_{{\rm H}}} \right) = \log\left( \frac{X_{{\rm Fe}}}{X_{{\rm H}}} \right)_{\odot}
                                           + \log \left(\frac{X'_{{\rm H}}}{X^{\odot}_{{\rm H}}} \right) - \Gamma_{Z}
\label{fe_gamma}
\end{equation}

\noindent
It is worth calling attention that at given total metallicity Z
different patterns of [$\rm X_{j}/X_{{\rm Fe}}$] may yield the same
total enhancement factor $\Gamma_{Z}$. This fact bears very much on
the procedure for deriving $\alpha$-enhanced indices because each
elemental species brings a different effect.

Finally, for the sake of comparison with the results from enhancing at
varying metallicity (see below), we enhance the elements as in
\citet{Munari05} by the ratio [$\alpha$/Fe]=+0.4\,dex and derive the
total enhancement factor $\Gamma_{Z}$ and the abundances listed in
Table~\ref{tab-enh}.

This definition of enhancement has been adopted by \citet{Tantalo98a},
\citet{Trager20a,Trager20b}, \citet{Maraston03}, \citet{Thomas03,Thomas03a},
\citet{ThoMara03} and TC04. It is indeed the most popular one to calculate
theoretical indices for SSPs.

\begin{table*}
\normalsize
\begin{center}
\caption[]{Abundance ratios for the solar-scaled and \alfe\ mixtures in which enhancement
is performed at constant total metallicity. The solar scaled values are taken from
\citet{GrevesseSauval98}. The enhancement factor \xsfe=+0.4\,dex is the same as in
\citet{CastelliKurucz04}.}
\label{tab-enh}
\small
\begin{tabular*}{95mm}{c| c c| c c c r}
\hline
\multicolumn{1}{c|}{} &
\multicolumn{2}{c|}{$\Gamma_{Z}=0$} &
\multicolumn{4}{c}{$\Gamma_{Z}=$0.25} \\
\hline
\multicolumn{1}{c|}{Element} &
\multicolumn{1}{c}{A$_{{\rm el}}$} &
\multicolumn{1}{c|}{X'$_{{\rm el}}$/Z} &
\multicolumn{1}{c}{[$X_{{\rm el}}/{{\rm Fe}}$]} &
\multicolumn{1}{c}{A$_{{\rm el}}$} &
\multicolumn{1}{c}{X'$_{{\rm el}}$/Z} &
\multicolumn{1}{c}{[$X_{{\rm el}}/{{\rm H}}$]} \\
\hline
C  & 8.52 & 0.1627 & 0.00 & 8.52 & 0.0840 &--0.2459 \\
N  & 7.92 & 0.0477 & 0.00 & 7.92 & 0.0246 &--0.2459 \\
O  & 8.83 & 0.4430 & 0.40 & 9.23 & 0.5747 &  0.1541 \\
Ne & 8.08 & 0.0985 & 0.40 & 8.48 & 0.1277 &  0.1541 \\
Na & 6.33 & 0.0020 & 0.00 & 6.33 & 0.0010 &--0.2459 \\
Mg & 7.58 & 0.0374 & 0.40 & 7.98 & 0.0485 &  0.1541 \\
Si & 7.55 & 0.0407 & 0.40 & 7.95 & 0.0528 &  0.1541 \\
S  & 7.33 & 0.0280 & 0.40 & 7.73 & 0.0363 &  0.1541 \\
Ca & 6.36 & 0.0037 & 0.40 & 6.76 & 0.0049 &  0.1541 \\
Ti & 5.02 & 0.0002 & 0.40 & 5.42 & 0.0003 &  0.1541 \\
Cr & 5.67 & 0.0010 & 0.00 & 5.67 & 0.0005 &--0.2459 \\
Fe & 7.50 & 0.0725 & 0.00 & 7.50 & 0.0374 &--0.2459 \\
Ni & 6.25 & 0.0043 & 0.00 & 6.25 & 0.0022 &--0.2459 \\
\hline
\end{tabular*}
\end{center}
\end{table*}

\subsubsection{Enhancement at varying metallicity}\label{enha_2}

Let us relax the condition that $X_H$, $X_{He}$ and Z remain
constant. By supposing that some elements are increased by the
factor $[X_i/X_{Fe}]>0$, the total metallicity will be greater, the
abundance of H and He will be different (decrease) and the condition
$\sum X_j=1$, where $j$ runs over all elements (starting from H),
will not be satisfied. In order to recover it, the new abundances
are re-scaled by means of the relations

\begin{equation}
N'_{j} = \frac{N_{j}}{\sum N_{j}} \quad {\rm } \quad X'_{j} =
\frac{X_{j}}{\sum X_{j}}.
\label{fe_gamma_new}
\end{equation}

\noindent
where once again $j$ runs over all elements. In this case $\Gamma_{Z}$
defined in Section~\ref{enha_1} can no longer be used to rank the
degree of total enhancement because it would vary with the
metallicity. We need indeed to will find a new, metallicity
independent, definition for the total enhancement factor. $\Gamma_{Z}$
can be simply replaced by the ratio long ago proposed by
\citet{Salaris93},

\begin{equation}
\Gamma_{\alpha}=\left[ \frac{\sum X_i^{{\rm \alpha,enh}}}{\sum X_j^{{\rm \alpha}}} \right]
\end{equation}

\noindent 
where $\sum X_i^{{\rm \alpha, enh}}$ is the sum of the abundances of
all enhanced elements and $\sum X_j^{{\rm \alpha}}$ is the sum of the
abundances of all $\alpha$-elements that are left unchanged. The
square brackets have their usual meaning.  $\Gamma_{\alpha}$ is
thereinafter use to indicate the global enhancement ratio. Given a set
of enhanced elements, $\Gamma_{\alpha}$ does not change with the
metallicity; it changes, however, with the specific pattern of
enhanced elements and degree of enhancement. To a certain extent the
same $\Gamma_{\alpha}$ may correspond to different patterns of
abundances\footnote{With the pattern of abundances we have adopted
$\Gamma_{\alpha}=0.4$ It also worth noticing that applying this new
definition of enhancement to the previous case we would also obtain
$\Gamma_{\alpha}\sim$0.4.}.

This definition of enhancement is for instance adopted by
\citet{CastelliKurucz04} on which the library of synthetic spectra by
\citet{Munari05} is based. The difference brought about by the two
definitions is not trivial and leads to different results.

\subsubsection{General remarks}

With the first definition (constant metallicity) indices for solar
scaled and $\alpha$-enhanced mixtures with the same Z can be simply
compared. With the second one for any degree of enhancement one has to
establish a priori the corresponding total metallicity before doing
the above comparison. For the sake of illustration and limited to the
case of solar composition [X=0.7347, Y=0.248, Z=0.0170], we present in
Table~\ref{tab-enh_zevar}, how the metallicity Z and abundance of H
and He would vary when the some heavy elements are enhanced by
+0.4\,dex.

The new abundances should be compared with those presented in
Table~\ref{tab-enh} obtained at constant Z. Due to the effect of
enhancement (the same on the same elements) the new chemical
parameters become [X=0.7219, Y=0.244, Z=0.0340]: the metallicity is
almost doubled and the abundances of H and He are slightly
decreased. Finally, in Table~\ref{enh_chem} for all metallicities
[M/H] of the \citet{Munari05} library we present the relationship
between the chemical parameters [X,Y,Z] for solar-scaled mixtures and
the corresponding ones for the enhancement [$\alpha$/Fe]=+0.4\,dex
($\Gamma_{\alpha}$=0.4) that are shortly indicated as [X',Y',Z'].

As in our study we have adopted the \citet{CastelliKurucz04} model
atmospheres and the companion \citet{Munari05} synthetic spectra we
also adopt their definition of enhancement. Throughout this paper,
when comparing results for the ``same'' metallicity and different
degree of enhancement we will make use of the relationship shown in
Table~\ref{enh_chem}. It is worth remarking that this relationship
changes with the adopted degree of enhancement. This is a point to
keep in mind when comparing results from different authors and also
results obtained with the definition of enhancement at constant
metallicity. Taking advantage from the fact that in the
\citet{Munari05} library the same enhancement factor [$\alpha/Fe$]
is adopted we will indicate our $\alpha$-enhanced mixtures simply
with $\Gamma_{\alpha}$=0.4.

\begin{table*}
\normalsize
\begin{center}
\caption[]{Abundance ratios for the solar-scaled and \alfe\ mixtures in which
enhancement is performed at varying the total metallicity Z. The
solar-scaled values are taken from \citet{GrevesseSauval98}. The
enhancement factor \xsfe=+0.4\,dex ($\Gamma_{\alpha}$=0.4) is the same
as in \citet{CastelliKurucz04}.}
\label{tab-enh_zevar}
\small
\begin{tabular*}{145mm}{c| c c c c| c c c c c }
\hline
\multicolumn{1}{c|}{} &
\multicolumn{4}{c|}{$\Gamma_{\alpha}$=0} &
\multicolumn{5}{c}{$\Gamma_{\alpha}$=0.4} \\
\hline
\multicolumn{1}{c|}{Element} &
\multicolumn{1}{c}{A$_{{\rm el}}$} &
\multicolumn{1}{c}{X$_{{\rm el}}$} &
\multicolumn{1}{c}{X$_{{\rm el}}/$Z} &
\multicolumn{1}{c|}{X$_{{\rm el}}/{{\rm H}}$} &
\multicolumn{1}{c}{[X$_{{\rm el}}/{{\rm Fe}}$]} &
\multicolumn{1}{c}{A$_{{\rm el}}$} &
\multicolumn{1}{c}{X$_{{\rm el}}$} &
\multicolumn{1}{c}{X$_{{\rm el}}/$Z} &
\multicolumn{1}{c}{X$_{{\rm el}}/{{\rm H}}$}\\
\hline
H  & 12.00 & 0.734709 &   --   & 1.0000 & 0.00 & 12.00 & 0.721943 &   --   & 1.0000 \\
He & 10.93 & 0.248336 &   --   & 0.3380 & 0.00 & 10.93 & 0.244022 &   --   & 0.3380 \\
C  & 8.52  & 0.002899 & 0.1709 & 0.0039 & 0.00 & 8.52  & 0.002849 & 0.0837 & 0.0039 \\
N  & 7.92  & 0.000849 & 0.0501 & 0.0012 & 0.00 & 7.92  & 0.000834 & 0.0245 & 0.0012 \\
O  & 8.83  & 0.007884 & 0.4650 & 0.0107 & 0.40 & 9.23  & 0.019460 & 0.5717 & 0.0270 \\
Ne & 8.08  & 0.001768 & 0.1043 & 0.0024 & 0.40 & 8.48  & 0.004365 & 0.1282 & 0.0060 \\
Na & 6.33  & 0.000036 & 0.0021 & 0.0000 & 0.00 & 6.33  & 0.000035 & 0.0010 & 0.0000 \\
Mg & 7.58  & 0.000674 & 0.0397 & 0.0009 & 0.40 & 7.98  & 0.001662 & 0.0488 & 0.0023 \\
Si & 7.55  & 0.000726 & 0.0428 & 0.0010 & 0.40 & 7.95  & 0.001793 & 0.0527 & 0.0025 \\
S  & 7.33  & 0.000500 & 0.0295 & 0.0007 & 0.40 & 7.73  & 0.001233 & 0.0362 & 0.0017 \\
Ca & 6.36  & 0.000067 & 0.0039 & 0.0001 & 0.40 & 6.76  & 0.000165 & 0.0048 & 0.0002 \\
Ti & 5.02  & 0.000004 & 0.0002 & 0.0000 & 0.40 & 5.42  & 0.000009 & 0.0003 & 0.0000 \\
Cr & 5.67  & 0.000018 & 0.0010 & 0.0000 & 0.00 & 5.67  & 0.000017 & 0.0005 & 0.0000 \\
Fe & 7.50  & 0.001287 & 0.0759 & 0.0018 & 0.00 & 7.50  & 0.001265 & 0.0372 & 0.0018 \\
Ni & 6.25  & 0.000076 & 0.0045 & 0.0001 & 0.00 & 6.25  & 0.000075 & 0.0022 & 0.0001 \\
\hline
\multicolumn{1}{c|}{}&
\multicolumn{4}{c|}{X=0.7347  Y=0.248  Z=0.0170} &
\multicolumn{5}{c} {X=0.7219  Y=0.244  Z=0.0340}\\
\hline
\end{tabular*}
\end{center}
\end{table*}

\begin{table*}
\normalsize
\begin{center}
\caption[]{Relationship between the chemical composition parameters [X, Y, Z] for
solar-scaled and $\alpha$-enhanced mixtures ([X', Y', Z']) according
to the definition of \citet{CastelliKurucz04}. This relationship holds
good only for the case under consideration. It changes indeed with
\asfe. The data refers to the mixtures in Table~\ref{tab-enh_zevar}.}
\label{enh_chem}
\small
\begin{tabular*}{105mm}{c c c r| c c c r}
\hline
\multicolumn{4}{c|}{$\Gamma_{\alpha}$=0} &
\multicolumn{4}{c}{$\Gamma_{\alpha}$=0.4} \\
\hline
\multicolumn{1}{c}{X} &
\multicolumn{1}{c}{Y} &
\multicolumn{1}{c}{Z} &
\multicolumn{1}{c|}{[M/H]} &
\multicolumn{1}{c}{X'} &
\multicolumn{1}{c}{Y'} &
\multicolumn{1}{c}{Z'} &
\multicolumn{1}{c}{[M/H]'}\\
\hline
0.7473 & 0.253 & 0.0002 & --2.0 & 0.7471 & 0.253 & 0.0004 & --1.7 \\
0.7470 & 0.252 & 0.0005 & --1.5 & 0.7465 & 0.252 & 0.0011 & --1.2 \\
0.7461 & 0.252 & 0.0017 & --1.0 & 0.7448 & 0.252 & 0.0035 & --0.7 \\
0.7433 & 0.251 & 0.0054 & --0.5 & 0.7391 & 0.250 & 0.0110 & --0.2 \\
0.7347 & 0.248 & 0.0170 &   0.0 & 0.7219 & 0.244 & 0.0340 &   0.3 \\
0.7087 & 0.240 & 0.0517 &   0.5 & 0.6725 & 0.227 & 0.1003 &   0.8 \\
\hline
\end{tabular*}
\end{center}
\end{table*}

\subsection{The \citet{Worthey94} and \citet{Borges95} fitting functions}\label{worthey_ff}

In the following we adopt the \citet{Worthey94} $\mathcal{FF}$s,
extended however to high temperature stars (\Teff$\approx$10000\,K) as
reported in \citet{Longhetti98a}. Recently new $\mathcal{FF}$s are
available also for the $\lambda$4000 break by \citet{Gorgas99} and Ca
II triplet by \citet{Cenarro01}. All these $\mathcal{FF}$s depend on
stellar effective temperature, gravity and metallicity ([Fe/H]). They
do not explicitly include the effect of $\alpha$-enhancement.

However, many studies have emphasized that absorption-line indices
should also depend on the detailed pattern of chemical abundances
\citep{Barbuy94,Idiart95,Weiss95,Borges95}, in particular when some
elements are enhanced with respect to the solar value. Empirical
$\mathcal{FF}$s for \mgii\ and \nad\ in which the effect of
enhancement is considered have been presented by \citet{Borges95}. For
all other indices one has to use the \citet{Worthey94}
$\mathcal{FF}$s, in which the effect of enhancement is missing, but
for the re-scaling of [Fe/H] given by equation~(\ref{fe_gamma}) and/or
(\ref{fe_gamma_new}). A great deal of the effect of enhancing
$\alpha$-elements is simply lost.

\section{The current correcting algorithm:
the \citet{Tripicco95} response functions}\label{tribell}

A general method designed to include the effects of enhancement on all
indices at once has been suggested by \citet[][]{Tripicco95}, who
introduced the concept of $\mathcal{RF}$s. In brief from model
atmospheres and spectra for three prototype stars, i.e. a cool-dwarf
star (CD) with \Teff=4575\,K and \logG=4.6, a turn-off (TO) star with
\Teff=6200\,K and \logG=4.1, and a cool-giant (CG) star with
\Teff=4255\,K and \logG=1.9, they calculate the reference indices
$I_{0}$ for the solar abundance ratios. Separately, doubling the
abundance $X_{i}$ of the C, N, O, Mg, Fe, Ca, Na, Si, Cr and Ti in
steps of $\Delta [X_{i}/{\rm H}]$=0.3\,dex, they determine the
variation $\Delta I = I_{{\rm enh}}-I_{0}$ in units of the
observational error $\sigma_{0}$. The indices $I_{0}$, the
observational error $\sigma_{0}$ and the normalized $\Delta I$ are
given in Tables~4, 5 and 6 of \citet{Tripicco95}.

The definition of the generic $\mathcal{RF}$ to be used for arbitrary
variations $\Delta [X_{i}/{\rm H}]$ is

\begin{displaymath}
{\rm R}_{0.3}(X_{i}) = \frac{1}{I_{0}}\, \frac{\Delta I}{\Delta [X_{i}/{\rm H}]}\, 0.3
\end{displaymath}

\noindent
The $\mathcal{RF}$s for cool-dwarfs, cool-giants and turn-off stars
constitute the milestones of the calibration. They are used by
\citet{Trager20a}, \citet{Thomas03} and \citet[][TC04]{Tantalo04a} to
transfer indices with solar abundance ratios to those enhanced in
$\alpha$-elements by means of two different algorithms. As they are
not strictly equivalent, some clarification is worth here.

Without providing a formal justification, \citet{Trager20a} propose
that the fractional variation of an index to changes of the chemical
parameters is the same as that for the reference index $I_{0}$
according to the relation

\begin{equation}
{\frac{\Delta I}{I}} = {\frac{\Delta I_{0}}{I_{0}}} =
 \left\{ \prod_{i} [1 + {\rm R}_{0.3}(X_{i})]^\frac{[X_{i}/{\rm H}]}{0.3} \right\} -1
\label{dind}
\end{equation}

\noindent 
where ${\rm R}_{0.3}(X_{i})$ are the $\mathcal{RF}$s we have defined
above. The explanation of relation~(\ref{dind}) has been provided by
TC04 to whom the reader should refer for details. The advantage with
this formulation is that no particular constraint is required on the
sign of $I_{0}$, and the $\Delta I$ given by
\citet{Tripicco95} are straightforwardly used.

A different reasoning has been followed by \citet{Thomas03}. In brief,
they start from the observational hint that in galactic stars
\mgii\,$\propto \exp({\rm [Mg/H]})$ \citep[see][]{Borges95}, assume that
all indices depend exponentially on the abundance ratios and introduce
the variable $\ln I \propto [X_{i}/{\rm H}]$, expand $\ln I$ in the
Taylor series and, by replacing the partial derivatives with respect
to abundances by the finite incremental ratio of \citet{Tripicco95},
express the fractional variation of an index to changes in the
abundance ratios as

\begin{equation}
\frac{\Delta I}{I} = \left\{ \prod_{i} exp \left( {\rm R}_{0.3}(X_{i}) \right)
               ^{\frac{[X_{i}/{\rm H}]}{0.3}} \right\} - 1
\label{dind_thomas}
\end{equation}

\noindent
The reader is referred to \citet{Thomas03} for the formal derivation
of relation~(\ref{dind_thomas}). Although relations~(\ref{dind}) and
(\ref{dind_thomas}) may look similar, actually they do not because in
the latter the partial derivatives in the Taylor expansion have been
replaced by the finite incremental ratios. This topic has been
thoroughly discussed by TC04.

The following remarks are worth to be made here. First, as emphasized
by \citet{Trager20a} and \citet{Thomas03}, equations~(\ref{dind}) and
(\ref{dind_thomas}) while securing that the indices tend to zero for
small abundances they let them increase with the exponent $[X_{i}/{\rm
H}]/0.3$. For abundances higher than $[X_{i}/{\rm H}]$=0.6\,dex, the
exponent may became too large and consequently both types of
correction may diverge. Second, the use of equations~(\ref{dind}) and
(\ref{dind_thomas}) requires that the sign of the index to be
corrected is the same of the corresponding $I_{0}$ in
\citet{Tripicco95}. In general this holds good. In principle, it may,
however, happen that the signs do not coincide. Furthermore, some of
the indices $I_{0}$ in \citet{Tripicco95} are negative so that the
$\ln I$ variable cannot be defined and the Taylor expansions can no
longer be applied. To overcome this potential difficulty,
\citet{Thomas03} apply a correcting procedure forcing the negative
reference indices $I_{0}$ of \citet{Tripicco95} to become
positive. This occurs in particular for \Hbeta\ of cool-dwarf stars
(and other indices as well). The argument is that neglecting non-LTE
effects \citet{Tripicco95} underestimate the true values of \Hbeta\ so
that negative values found for temperatures lower than approximately
4500\,K should be shifted to higher, positive values \citep[see for
instance Fig.~12 in][]{Tripicco95}. We suspect that any change to the
values tabulated by \citet{Tripicco95} may be risky for a number of
reasons: (i) The $\mathcal{FF}$s have been derived from a set of data
that include a significant number of stars with negative values of
\Hbeta; (ii) the incremental ratios by \citet{Tripicco95} have been
calculated for particular stars (stellar spectra) with assigned \Teff,
\logG\ and $I_{0}$. Changing the reference $I_{0}$ while leaving
unchanged the incremental ratios (partial derivatives) may not be very
safe. (iii) The replacement of the partial derivatives with the
\citet{Tripicco95} incremental ratios may be risky when dealing with
exponential functions; (iv) the corrections found by \citet{Thomas03}
for a number of indices are quite large; (v) finally, the use of $\ln
I$ as dependent variable which requires that only positive values for
$I_{0}$ are considered.

The application of eqns.~(\ref{dind}) and (\ref{dind_thomas}) by TC04
has generated different and highly controversial results for some
indices, \Hbeta\ in particular, under unusual enhancements in some
elements. In brief, TC04 presented grids of indices with
$\alpha$-enhanced abundance ratios. The indices were derived from the
\citet{Salasnich20} stellar models and isochrones with the abundance
ratios by \citet{Ryan91}, and corrected by means of
equation~(\ref{dind}) of \citet{Trager20a}. The unusual abundance
ratio for Ti ([Ti/Fe]=0.63, which implies [Ti/H]=0.2634 when the
definition of enhancement at constant Z is adopted), together with the
interpolation among the fractional variations of the calibrating
stars, yielded \Hbeta\ strongly increasing with $\Gamma_{Z}$. This
immediately reflected on the ages assigned to galaxies by means of the
minimum-distance method of \citet{Trager20a,Trager20b} widely used by
TC04. The strong impact of Ti on \Hbeta\ is due to the high
$\mathcal{RF}$s of this element for cool-dwarfs in the
\citet{Tripicco95} calibration.  Decreasing the ratio [Ti/Fe] to
zero as in \citet{Trager20a} or to 0.3 as in \citet{Thomas03}, the
results by TC04 were in close agreement with those by the other
authors. It is worth recalling that the ratio [Ti/H] entering
equations~(\ref{dind}) and (\ref{dind_thomas})\footnote{Similar
considerations apply to all other elements listed in the Tripicco \&
Bell tabulation.}, was 0 in \citet{Trager20a}, 0.023 in
\citet{Thomas03} and nearly 0.2 in TC04 with obvious consequences.
The effect was also confirmed by adopting the more recent
determinations by \citet{Carney96} and \citet{Habgood01} of the
[Ti/Fe] in globular clusters that yield
$\langle$[Ti/Fe]$\rangle$$\simeq$0.25$\div$0.30, and by
\citet{Gratton03} for a sample of metal-poor stars with accurate
parallaxes for which they yields
$\langle$[Ti/Fe]$\rangle$$\simeq$0.20$\pm$0.05 (TC04). In any case the
controversy about the correcting technique still remained. It can be
reduced to the statement: does \Hbeta\ increase (significantly) with
\asfe\ or not? Or, even worse, does it decrease with it? According to
\citet{Thomas03} there should be no difference in \Hbeta\ at
increasing \asfe. The analysis below will clarify that this is not the
case.

\section{The new response functions}\label{new_calib}

In this section we present the $\mathcal{RF}$s we derive from high
resolution spectra and a simple algorithm to pass from solar-scaled to
$\alpha$-enhanced indices.

\subsection{The high-resolution spectra}\label{high_res_sp}

The spectra used in the present analysis are taken for a partial
pre-release of the 2500--10500\AA\ extensive synthetic spectral
library computed by \citet{Munari05}. The spectral library is based on
the new grid of ATLAS9 model atmospheres computed by
\citet{CastelliKurucz04} for new opacity distribution functions
that include, among other characteristics, the replacement of the
solar abundances by \citet{Anders89} with those from
\citet{GrevesseSauval98} and the TiO line-list by \citet{Kurucz93}
with that of \citet{Schwenke98}. The \citet{Munari05} spectral library
includes more than 200,000 spectra at resolutions of 20,000 and 2000
(the latter matching the SLOAN spectra), as well as uniform
dispersions of 1\AA/pix and 10\AA/pix.

The library covers very large intervals of effective temperature
(3500--50000\,K), gravities (0$\leq$\logG$\leq$5.0), metallicities
(--2.5$\leq$${\rm [Z/Z_{\odot}]}$$\leq$0.5), microturbulent velocities
(0, 1, 2 and 4 ${\rm km/sec}$), a dozen of rotational velocities, and
finally two degrees of enhancement (\asfe=0. and +0.4\,dex). For all
other details see \citet{Munari05}. The library is still under
construction. When completed, it will provide a powerful tool for
predicting absorption line indices and exploring their applications. A
strictly coordinated library of synthetic spectra is the one published
by \citet{Zwitter04} for the GAIA and RAVE wavelength range, amounting
to 183,588 spectra covering a similar space of parameters.

In this work we have made use of the version at 1\AA/pix with
temperature interval 4000--13000\,K, gravity 0$\leq$\logG$\leq$5.0,
metallicities --2.5$\leq$${\rm [Z/Z_{\odot}]}$$\leq$0.5, null
micro-turbulence and rotational velocities and the two degrees of
enhancement (\asfe=0. and +0.4\,dex, i.e. $\Gamma_{\alpha}$=0. and
0.4, respectively). It is worth recalling that in this latter case the
temperature coverage is narrower at high metallicities, i.e.  from
4000 to 7250\,K for $[Z/Z_{\odot}]$=0 and from 10000 to 13000 for
$[Z/Z_{\odot}]$=0.5.

\begin{figure}
\psfig{file=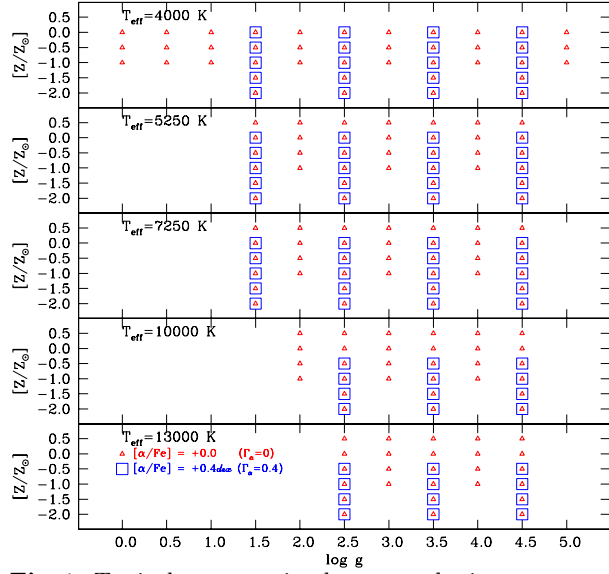,width=8.0truecm}
   \caption{Typical coverage in the atmospheric parameters, and \gaa\ of the
   1-\AA\ resolution spectra used in this study. The filled-triangles
   are for chemical compositions with solar abundance ratios \asfe=0
   (\gaa=0), whereas the open-squares are $\alpha$-enhanced mixtures
   with \asfe=+0.4\,dex (\gaa=0.4). }
\label{grid}
\end{figure}

\begin{figure*}[ht]
\centerline{
\psfig{file=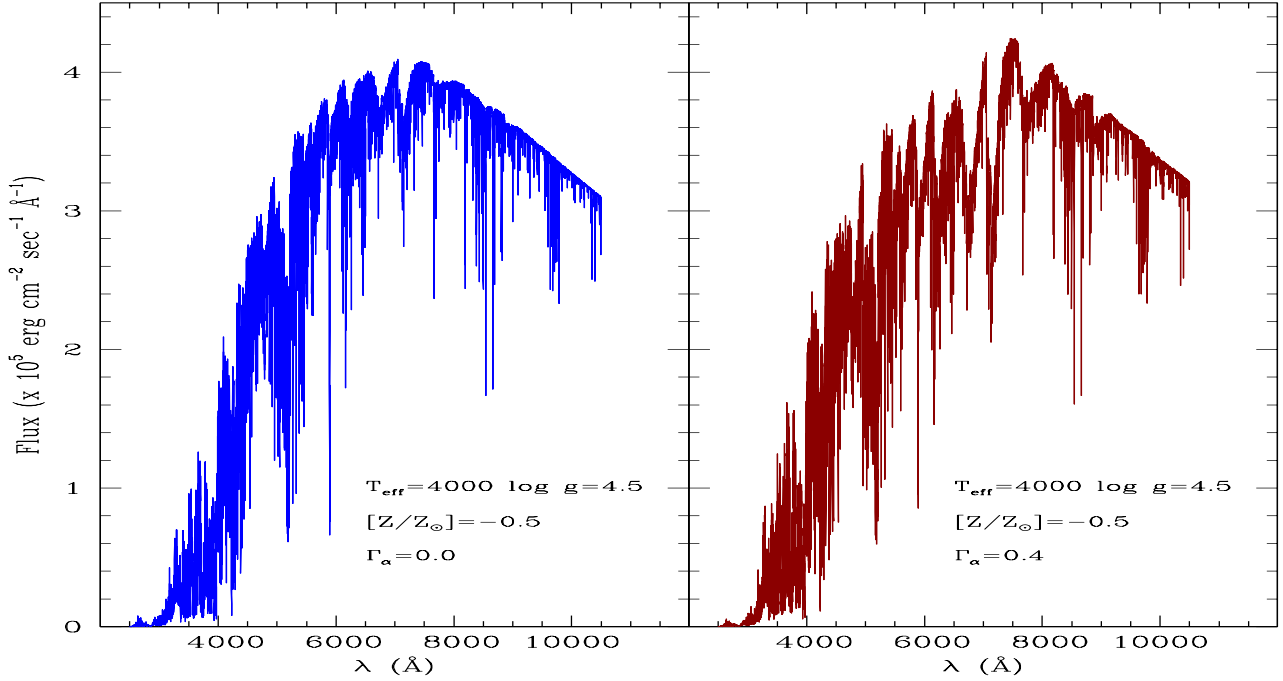,width=17.0truecm,height=9.0truecm}}
   \caption{Two theoretical spectra of the \citet{Munari05} library.
   Left-hand panel: spectral energy distribution
   F$_{\lambda}$ for the star with \Teff=4000\,K, \logG=4.5, ${\rm
   [Z/Z_{\odot}]}$=--0.5, \gaa=0. The wavelength is in
   \AA. F$_{\lambda}$ is in ${\rm erg~sec^{-1} cm^{-2} \AA^{-1}}$ and
   the resolution is 1\AA. Right-hand panel: the same but for
   \gaa=0.4.}
\label{spectrum_0_04}
\end{figure*}

\begin{figure*}
\centerline{
    \psfig{file=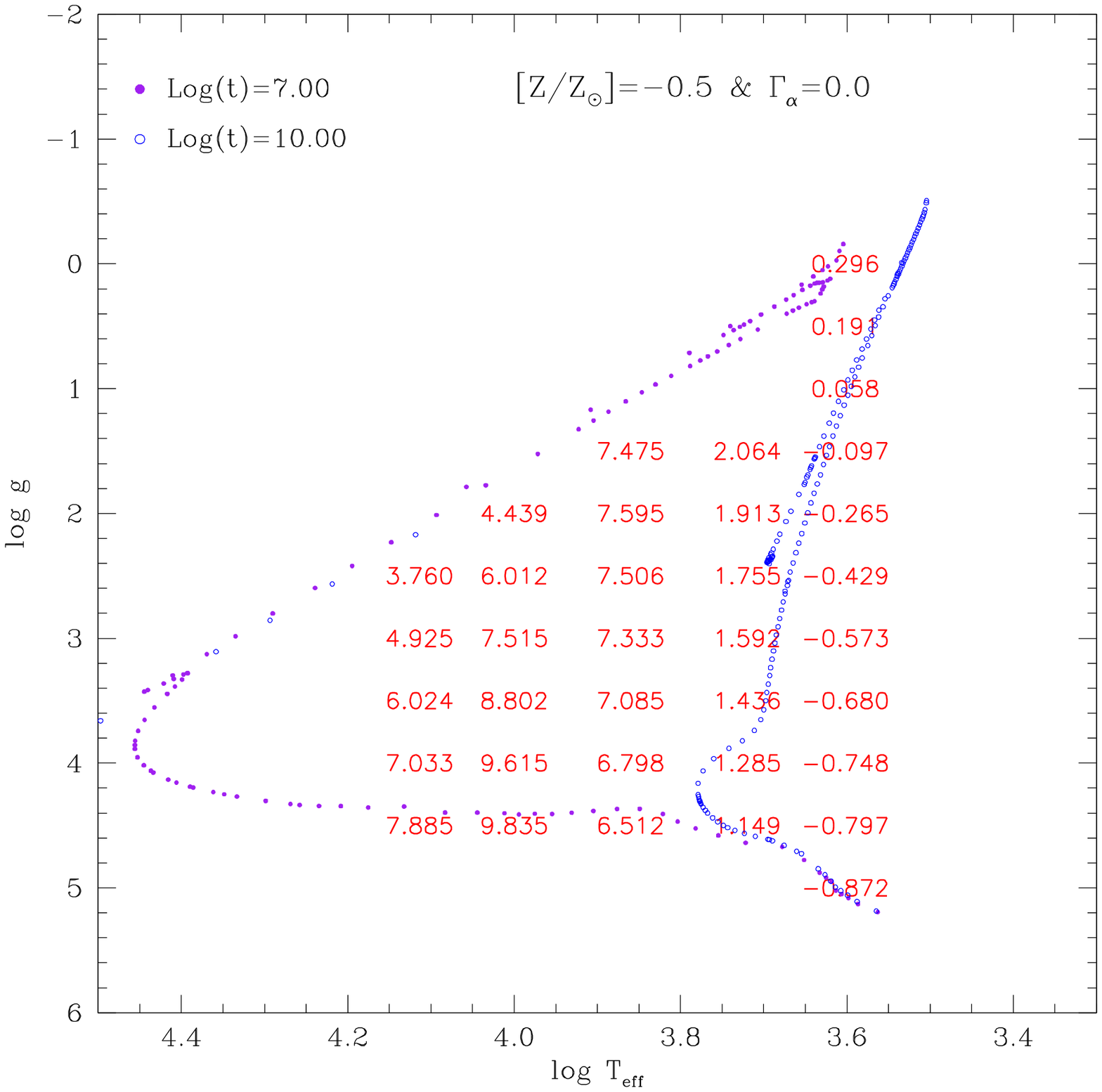,width=8.0truecm}\psfig{file=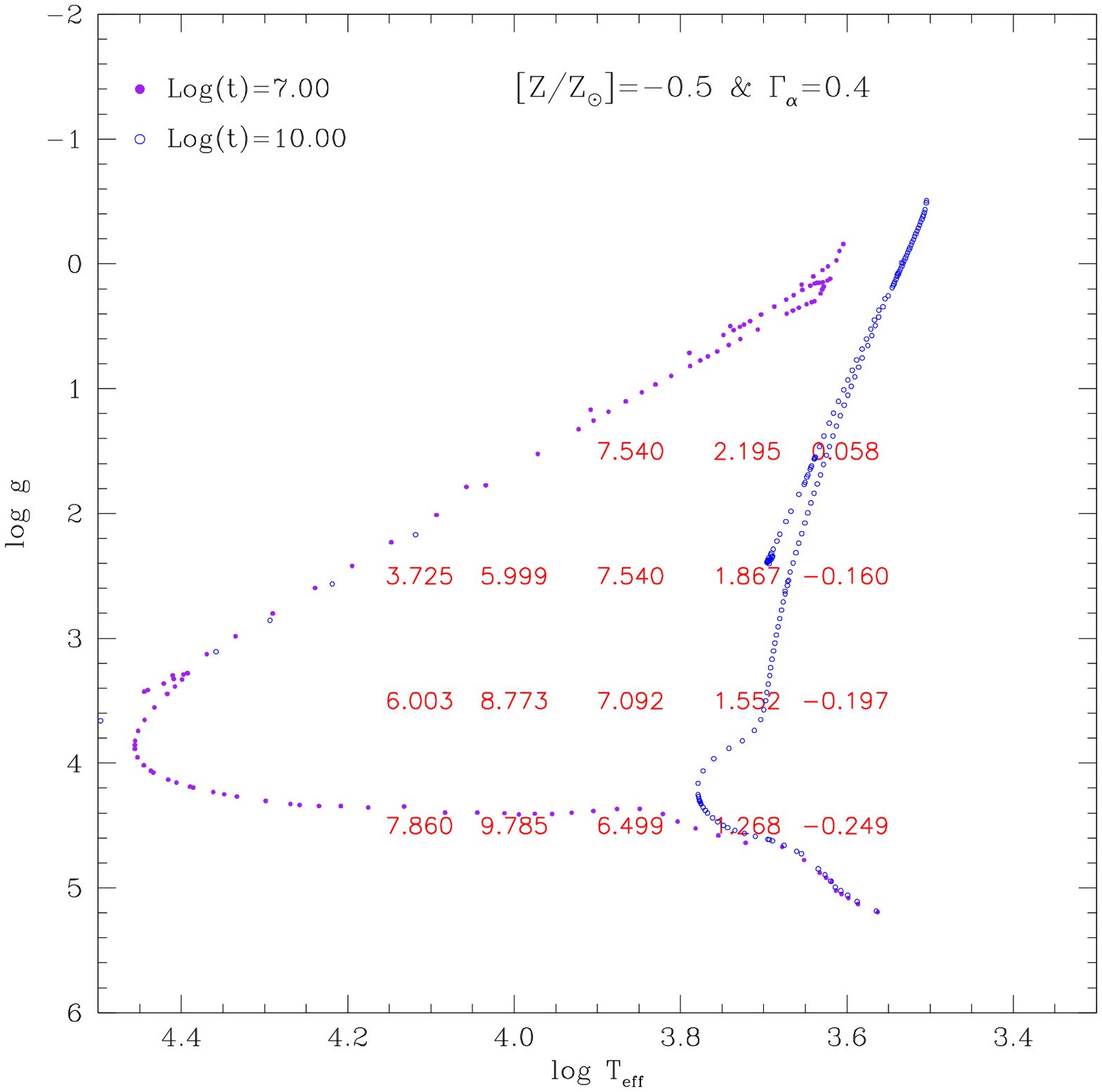,width=8.0truecm}
    } \caption{Left Panel Variation of \Hbeta\ from 1-\AA\ resolution
    spectra across the HR-diagram in the \logG\ versus \logT\
    plane. Each value of \Hbeta\ corresponds to a star (theoretical
    spectrum) of given \logG, \logT\ and ${\rm [Z/Z_{\odot}]}$. The
    case under consideration is for ${\rm [Z/Z_{\odot}]}$=--0.5 and
    \gaa=0. In the same diagram we also plot two isochrones of the
    Padova Library with the same metallicity (Z=0.008) and ages of
    0.01 and 10\,Gyr. {Right Panel: The same as in the left panel but
    for \gaa=0.4.}}
\label{ind_008_sol_enh}
\end{figure*}

The temperature, gravity and metallicity coverage of the body of
stellar spectra in usage here is summarized in Fig.~\ref{grid} (as
already mentioned in the case of the highest metallicity and
$\alpha$-enhanced spectra, the temperature interval is smaller). Each
panel is for a different effective temperature as indicated. The
symbols show the combination of metallicity, expressed here in
spectroscopic notation as ${\rm [Z/Z_{\odot}]}=log({\rm
Z/Z_{\odot}})$, and gravity (\logG) for which a spectrum has been
calculated. The triangles are for $\Gamma_{\alpha}$=0 whereas the
squares are for $\Gamma_{\alpha}$=0.4. The parameters span the
following ranges: two values of \asfe\ (or \gaa), i.e. \asfe=0 and
+0.4\,dex; six values of ${\rm [Z/Z_{\odot}]}$, i.e. --2.0, --1.5,
--1.0, --0.5, 0 and 0.5 for \asfe=0, and five values (${\rm
[Z/Z_{\odot}]}$=--2.0, --1.5, --1.0, --0.5 and 0) for \asfe=+0.4\,dex;
five values of \Teff, i.e.  4000, 5250, 7250, 10000 and 13000\,K for
most of the spectra, they are reduced to three values (4000, 5250 and
7250 K) for \asfe=+0.4\,dex and ${\rm [Z/Z_{\odot}]}$=0; finally
eleven values of \logG\ going from 0 to 5.0 in steps 0.5. The spectral
grid is only a sub-set of the original model atmosphere grid, however
fully adequate to the purposes of this study. The range of wavelength
in each spectrum goes from 2500\AA\ to 10500\AA\ with resolution of
1\AA. The flux F$_{\lambda}$ is in ${\rm erg~sec^{-1}cm^{-2}\AA^{-1}}$. 
In Fig.~\ref{spectrum_0_04} we show for the sake of illustration the
spectral energy distribution for the model with \Teff=4000\,K,
\logG=4.5, ${\rm [Z/Z_{\odot}]}$=--0.5, micro-turbulence velocity
$k$=2\,km/sec, and \gaa=0 (left panel) and 0.4 (right panel). It
fairly represents the case of a cool-dwarf star.

\subsection{Indices from 1-\AA\ resolution spectra}\label{ind_high_res}

With the aid of equations~(\ref{ind_def_ew}) and (\ref{ind_def_mag})
and the pass-bands of \citet{Trager98} we filter the energy
distribution of each star in our grids and derive the indices. They
are not strictly equivalent to the Lick system because the spectra in
usage have 1-\AA\ resolution, whereas those on the Lick system have
significantly smaller resolution (approximately 8.4\AA) that also
depends on the wavelength interval \citep[see][for all
details]{Ottaviani97}. This case will be considered in detail in
Section~\ref{degrade_to_lick}.

The data for the whole sets of indices are not displayed but they are
partly given in and partly recovered from Tables~A.1 through
A.5 to be described in Appendix A. We limit ourselves to
show in Figs.~\ref{ind_008_sol_enh} the variation of the index \Hbeta\
across the HR-diagram for the case with ${\rm [Z/Z_{\odot}]}$=--0.5
and \gaa=0 and 0.4. In the same diagram we also plot two isochrones
with the same metal content taken from the Padova Library (Girardi
private communication): the ages are 0.01\,Gyr and 10\,Gyr. The grids
of calibrating stars cover most of the HR-diagram in which real stars
are found.

\begin{figure}
\psfig{file=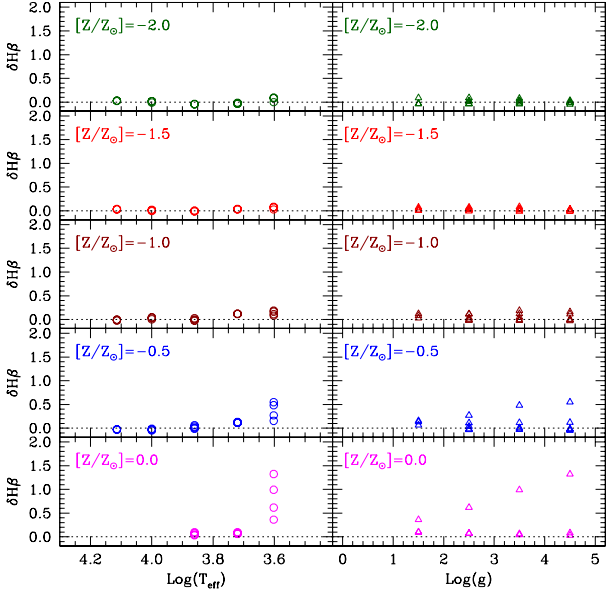,width=8.0truecm}
   \caption{1-\AA\ resolution spectra. Variations of the index \Hbeta\
   passing from solar to $\alpha$-enhanced chemical mixture. We plot
   the difference $\left( \delta I \right)_{\Teff,g,{\rm
   [Z/Z_{\odot}]}}$ as a function of the atmospheric parameters as
   indicated. Left-hand panels: for each \logT\ $\delta$H$\beta$
   increases with \logG. Right-hand panels: for each \logG\
   $\delta$H$\beta$ decreases with \logT.}
\label{delta_ind}
\end{figure}

\subsection{Response functions for 1-\AA\ resolution spectra}\label{tb95_like}

It is worth of interest here to present the analogue of the
\citet{Tripicco95} $\mathcal{RF}$s, the only major difference is that
they cannot be evaluated for separate increases in the abundance of
individual species [$\alpha_{i}/{\rm Fe}$] but only for all elements
enhanced with respect to the solar value lumped together. Now the
dependence of the $\mathcal{RF}$s on temperature, gravity, metallicity
and enhancement can be evaluated in detail. This was not the case with
the \citet{Tripicco95} calibration which stood on three stars with
solar metallicity. We start calculating the differences

\begin{equation}
\left(\delta I \right)_{\Teff,g,{\rm [Z/Z_{\odot}]}} = I_{{\rm enh}} - I_{{\rm sol}}
\end{equation}

\noindent
taken at fixed \logT, \logG\ and metallicity (${\rm [Z/Z_{\odot}]}$)
and varying \gaa\ from 0 to 0.4, whose corresponding indices are
indicated as $I_{{\rm sol}}$ and $I_{{\rm enh}}$, respectively. The
differences for the whole sets of values are given in
Tables~A.1--A.5 for ${\rm [Z/Z_{\odot}]}$=--2.0,
--1.5, --1.0, --0.5 and 0, respectively.

In the ten panels of Fig.~\ref{delta_ind} we show the case of
\Hbeta. Each panel displays the variation of \Hbeta\ at changing
\logT\ (left panels) and/or \logG\ (right panels). We note that, for 
any metallicity and passing from \gaa=0 to 0.4: (i) at given \logT\
$\delta$\Hbeta\ increases at increasing \logG; (ii) at given \logG\
$\delta$\Hbeta\ increases at decreasing \logT. For most cases
$\delta$\Hbeta\ is positive, i.e. the index \Hbeta\ increases at
increasing \gaa\ or, in other words, \Hbeta\ for $\alpha$-enhanced
mixtures is greater than the solar case, keeping all other parameters
constant. The increase is always significant, say approximately
0.1$\div$0.2, but it can amount to nearly 1.4 for stars of low \Teff\
and high gravity, roughly in the intervals covered by cool-dwarf and
old turn-off stars, which are the most interesting in view of the
forthcoming applications.

\begin{figure}
\psfig{file=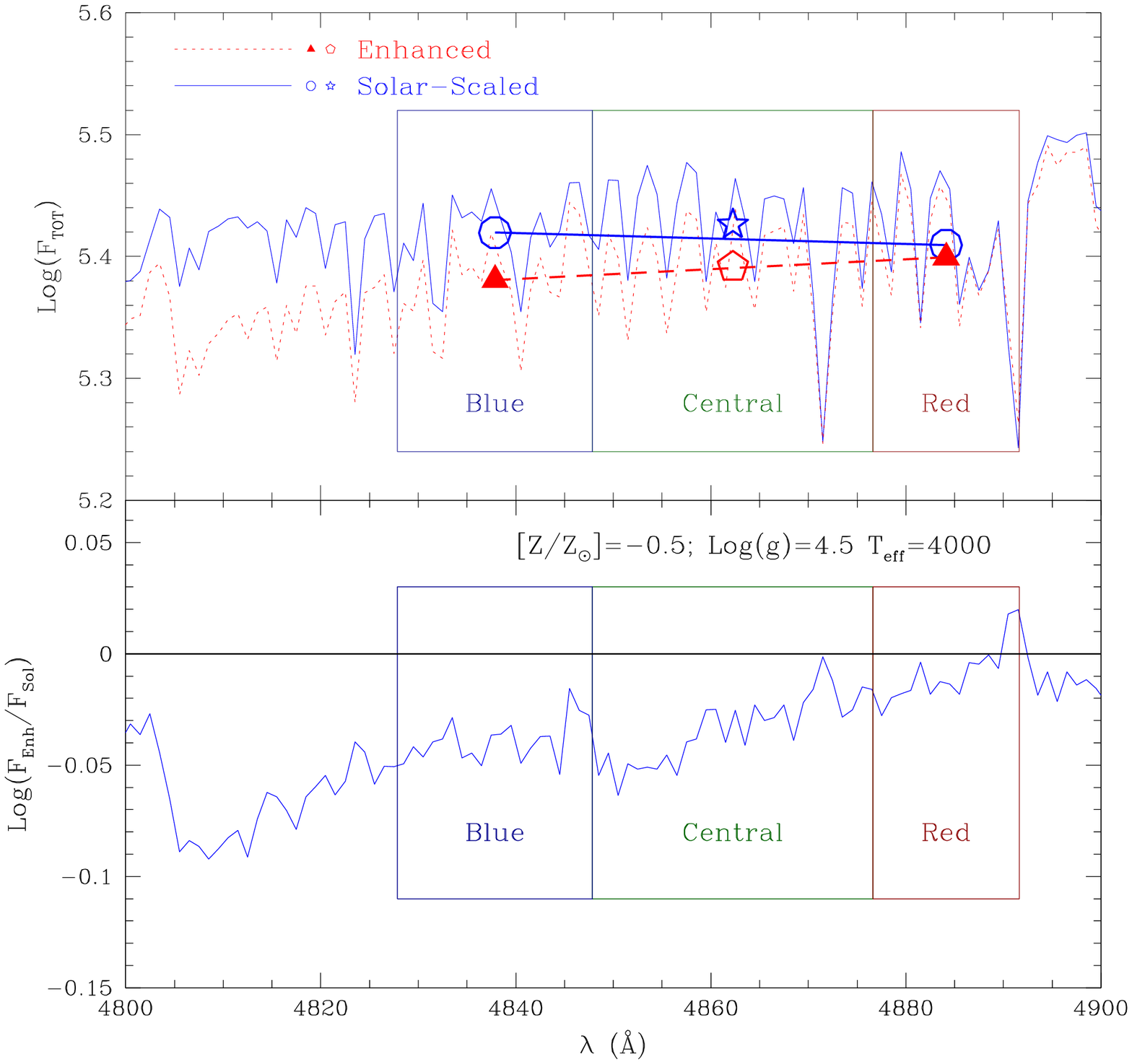,width=8.0truecm,height=9truecm}
   \caption{1-\AA\ resolution spectra. Building-up of the index
   \Hbeta\ in the cool-dwarf with \Teff=4000, \logG=4.5 and ${\rm
   [Z/Z_{\odot}]}$=--0.5 both for the solar scaled and the total
   enhancement factor \gaa=0.4. The bottom panel shows the ratio
   F$_{\lambda,{\rm enh}}/$F$_{\lambda,\odot}$. The absorption in the
   $\alpha$-enhanced spectrum is significantly larger than the
   solar-scaled one. The effect is larger in the blue wing of the
   pseudo-continuum and central band than in the red
   pseudo-continuum. The upper panel shows the spectral energy
   distribution of the solar-scaled (solid line) and $\alpha$-enhanced
   (dotted line) spectrum and the pass-bands defining the index
   \Hbeta. The open-circles and star show the mean fluxes in the three
   pass-band and the interpolation of the pseudo-continuum to derive
   F$_{c}$ in the case of solar scaled spectrum. The filled-triangles
   and pentagon are the same but for the $\alpha$-enhanced
   mixture. The increase of \Hbeta\ passing from solar to
   $\alpha$-enhanced abundance ratios is straightforward.}
\label{diff_in_hb}
\end{figure}

Why such an increase with $\alpha$? The answer lies in relative
variation of F$_{l}$ and F$_{c}$. In the case of \Hbeta\ (our
prototype index) we derive the variation $\left(\delta I\right)$
passing from solar to $\alpha$-enhanced mixtures. Upon differentiating
the index with respect to F$_{l}$ and F$_{c}$ ($\frac{\partial
I}{\partial {\rm F}_{l}}$=--$\frac{1}{{\rm F}_{c}}\Delta \lambda$ and
$\frac{\partial I}{\partial {\rm F}_{c}}$=$\frac{{\rm F}_{l}}{{\rm
F}_{c}^{2}} \Delta \lambda $) we obtain

\begin{equation}
\left(\delta I\right) = \Delta \lambda \left( \frac{{\rm F}_{l}}{{\rm F}_{c}} \right)_{{\rm sol}}
                        \left[ {\Delta {\rm ln~F}_{c}} - {\Delta {\rm ln~F}_{l}} \right]
\end{equation}

\noindent
where $\Delta {\rm ln~F}_{l} $ and $\Delta {\rm ln~F}_{c}$ are the
differences in the line and continuum fluxes passing from enhanced to
solar. Plugging the values of F$_{l}$ and F$_{c}$ as appropriate, it
turns out that the index increases with $\alpha$. What happens is best
illustrated in Fig.~\ref{diff_in_hb} which, limited to the case of a
typical cool-dwarf with \Teff=4000, \logG=4.5 and ${\rm
[Z/Z_{\odot}]}$=--0.5 (Z=0.008), shows how the index \Hbeta\ is built
up and how it varies passing from \gaa=0 to \gaa=0.4. First in the
bottom panel we display the ratio F$_{\lambda,{\rm enh}}/{\rm
F}_{\lambda,\odot}$ and the pass-bands defining the index
\Hbeta. The absorption in the $\alpha$-enhanced spectrum is
significantly larger than in the solar-scaled one. The effect is
larger in the blue pseudo-continuum and central band than in the red
pseudo-continuum. This means that $\alpha$-enhanced mixtures distort
the spectrum in such a way that simple predictions cannot be
made. This is due to the contribution of hundreds of molecular and
atomic lines falling into the spectral regions over which the index is
defined. Secondly, in the upper panel we show the spectral energy
distribution of the solar-scaled (solid line) and $\alpha$-enhanced
(dotted line) spectrum and once more the pass-bands for \Hbeta. The
open-circles and the big empty star show the mean fluxes in the three
pass-bands and the interpolation of the pseudo-continuum to derive
F$_{c}$ in the case of solar-scaled spectrum. The filled-triangles and
the empty pentagon are the same but for the $\alpha$-enhanced
mixture. The increase of \Hbeta\ passing from solar to
$\alpha$-enhanced abundance ratios is straightforward.

With the differences $\left(\delta I \right)$, the equivalent of the
\citet{Tripicco95} $\mathcal{RF}$s [i.e. ${\rm R}_{0.3}(X_{i})$s] is

\begin{equation}
{\rm R}_{0.4}(\alpha) =  \frac{1}{I_{{\rm sol}}}\,\frac{I_{{\rm enh}}-I_{{\rm sol}}}{\Delta[\alpha/{\rm Fe}]}\,0.4
\label{r04}
\end{equation}

\noindent
where the symbol $\alpha$ reminds the reader that only variations for
the all elements enhanced at a time are available (the products in
equations~(\ref{dind}) and (\ref{dind_thomas}) would extend over one
term only). Work is in progress to exactly repeat the analysis made by
\citet{Tripicco95}, i.e. to provide partial $\mathcal{RF}$s by separately
enhancing individual elements at a time. This would improve upon the
$\mathcal{RF}$s and yet maintain alive the $\mathcal{FF}$s until
high-resolution spectra will become a general tool.

\begin{figure}
\psfig{file=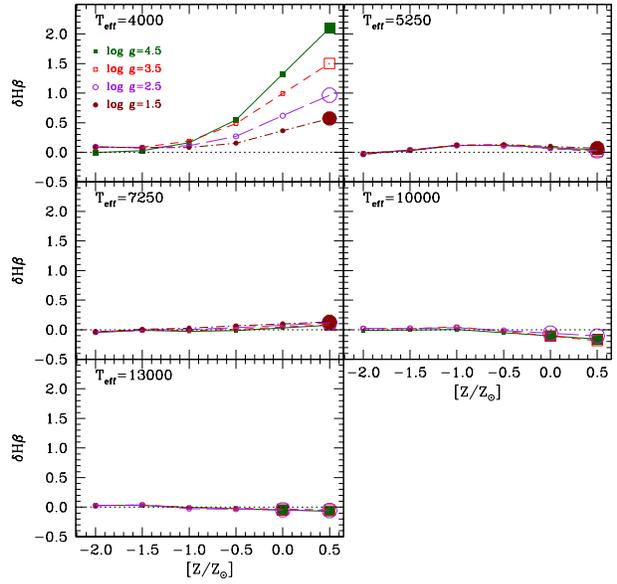,width=8.0truecm}
   \caption{1-\AA\ resolution spectra. The variation $\delta \Hbeta$
   as a function of the metallicity ${\rm [Z/Z_{\odot}]}$ at fixed
   \logT\ and \logG\ as indicated. In each panel the big symbols
   display the value of $\delta \Hbeta$ linearly extrapolated to the
   metallicity ${\rm [Z/Z_{\odot}]}$=0  for T$\leq$7250\,K and up
   to 0.5  for T=1000 and 13000\,K.}
\label{interp_z}
\end{figure}

Having done that, for the sake of a preliminary investigation of the
whole subject, we linearly extrapolate the results obtained for ${\rm
[Z/Z_{\odot}]}$=--2.0, --1.5, --1.0, --0.5 and 0 to the higher value
of metallicity, i.e. ${\rm [Z/Z_{\odot}]}$=0.5 for the temperatures
lower than 10000\,K and also to ${\rm [Z/Z_{\odot}]}$=0 and 0.5 for
the temperatures of 10000 and 13000\,K.

The extrapolation is safe, because $\left( \delta I \right)$=$I_{{\rm
enh}}$--$I_{{\rm sol}}$ smoothly change with the metallicity at given
\logT\ and \logG. Limited to the case of \Hbeta, this is shown in
Fig.~\ref{interp_z}. The dependence is almost linear and nearly
insensitive to gravity for \Teff=5250, 7250, 10000 and 13000\,K. The
case of \Teff=4000 deserves little attention, because $\delta \Hbeta$
tends to increase with ${\rm [Z/Z_{\odot}]}$ and gravity.  Adopting
the simple linear extrapolation on ${\rm [Z/Z_{\odot}]}$, we probably
underestimate $\delta \Hbeta$ for the high metallicities.

\subsection{Indices and response functions at the Lick resolution}\label{degrade_to_lick}

\begin{figure}
\psfig{file=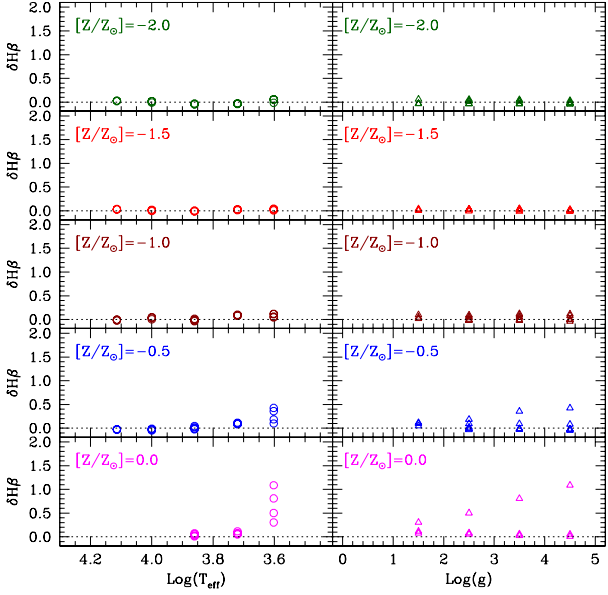,width=8.0truecm}
   \caption{The same as in Fig.~\ref{delta_ind} but for spectra
   degraded to the resolution of the Lick system. The details are
   given in Section~\ref{degrade_to_lick}. In the various panels we
   show the variations of the index \Hbeta\ passing from solar to
   $\alpha$-enhanced chemical mixture. We plot the difference $\left(
   \delta I \right)_{\Teff,g,{\rm [Z/Z_{\odot}]}}$ as a function
   of the atmospheric parameters as indicated. Left-hand panels: for
   each \Teff\ $\delta$H$\beta$ increases with \logG. Right-hand
   panels: for each \logG\ $\delta$H$\beta$ decreases with \logT.}
\label{delta_ind_bis}
\end{figure}

The 1-\AA\ resolution spectra have been degraded to the resolution of
the Lick system following the procedure and the FWHM of
\citet{Ottaviani97}. The new spectral energy distribution of each star
in the calibrating grid have been filtered to derive the indices and
the whole procedure described in Section~\ref{tb95_like} has been
repeated. The data are given in Tables~A.6--A.10 of Appendix A. These
are the analogue of Tables~A.1 through A.5. For the sake of comparison
we show in Fig.~\ref{delta_ind_bis} the differences $\left(\delta I
\right)_{\Teff,g,{\rm [Z/Z_{\odot}]}} = I_{{\rm enh}} -I_{{\rm
sol}}$. In general, the new $\left(\delta I
\right)_{\Teff,g,{\rm [Z/Z_{\odot}]}}$ share the same trends as in the
previous case, been however only slightly smaller. The most noticeable
change is for \Hbeta, whose largest $\left(\delta I \right)$ amounts
to 1.1 instead of 1.4 for the case of a cool-dwarf (\Teff=4000,
\logG=4.5) and ${\rm [Z/Z_{\odot}]}$=0. Also in this case we linearly
extrapolate the data to higher values of the metallicity for the sake
of preliminary investigation. Indices and $\left(\delta I \right)$
fully consistent with the Lick system will be used in the discussion
below.

\section{SSP indices from $\mathcal{FF}_{{\bf S}}$ and new $\mathcal{RF}_{{\bf S}}$}\label{ind_ssp}

Owing to the large body of data and synthetic indices on the Lick
system, it might be worth of interest to derive indices still based on
the $\mathcal{FF}$s but in which the old $\mathcal{RF}$s are replaced
by the new ones. This means that indices for solar scaled mixture are
calculated as amply described in TC04, whereas those for \gaa$>$0 are
obtained according to the following equation

\begin{equation}
I_{i,{\rm enh}}=I_{i,{\rm sol}} + \delta \mathcal{I}_{i}
\label{ind_diff_enha}
\end{equation}

\noindent 
where $I_{i,{\rm sol}}$ is the solar-scaled index of the generic star
in the SSP and $I_{i,{\rm enh}}$ the is same but corrected for
enhancement by $\delta \mathcal{I}_{i}$. This is derived by linearly
interpolating both in \logG\ and \logT\ the new $\mathcal{RF}$s
(listed in Tables~A.6--A.10) and taking into account the
enhancement-metallicity relationship discussed in
Section~\ref{enha}\footnote{It is worth recalling that an equivalent
procedure would be to use eqn. (\ref{dind}) combined with (\ref{r04})
instead of eqn. (\ref{ind_diff_enha}).}.

\subsection{Definition of SSP indices}\label{ind_ssp_def}

The integrated indices of SSPs can be derived in the following way. We
start from the flux in the absorption-line of the generic star of the
SSP, F$_{l,i}$

\begin{eqnarray}
{\rm F}_{l,i} &=& {\rm F}_{c,i} \left( 1 - \frac{I_{l,i}}
               {\Delta \lambda} \right) \,\,\, (EW)\\
{\rm F}_{l,i} &=& {\rm F}_{c,i} 10^{-0.4 I_{l,i}}\,\,\, (Mag)
\end{eqnarray}

\noindent 
where $I_{l,i}$ is the index derived from the $\mathcal{FF}$s using
the \logT, \logG\ and chemical composition of the star, and, in the
case of $\alpha$-enhancement, also corrected according to
equation~(\ref{ind_diff_enha}). F$_{c,i}$ is the pseudo-continuum
flux, F$_{l,i}$ is the flux in the pass-band and $\Delta \lambda$ is
the same as in equation~(\ref{ind_def_ew}). The flux F$_{c,i}$ is
calculated by interpolating to the central wavelength of the
absorption-line, the fluxes in the midpoints of the red and blue
pseudo-continua bracketing the line \citep{Worthey94}.

To calculate the flux F$_{c,i}$ of a generic star one needs the
theoretical spectrum of the same star. To be able to compare our
results based on the new $\mathcal{RF}$s with the old ones by
\citet{Trager20a}, \citet{Thomas03} and TC04 based on low-resolution
spectra, we have to make use of stellar spectra with similar
resolution. Therefore we leave aside the library of high-resolution
spectra of \citet{Munari05} and adopt the low-resolution one
amalgamated by \citet{Girardi02} and adopted by TC04. As the backbone
of this library are the Kurucz ATLAS9 model atmospheres and stellar
spectra, there is partial consistency between the spectra adopted to
evaluate the pseudo-continuum and the 1-\AA\ resolution spectra
adopted to derive the new $\mathcal{RF}$s.

Known the fluxes F$_{l,i}$ and F$_{c,i}$ for a single star, we weight
its contribution to the integrated value on the relative number of
stars of the same type. Therefore the integrated index is given by

\begin{eqnarray}
I_{l}^{{\rm SSP}} &=& \Delta \lambda
               \left( 1 - \frac{\sum_{i}{\rm F}_{l,i} {\rm N}_{i}}
               {\sum_{i} {\rm F}_{c,i} {\rm N}_{i}} \right)\,\,\, (EW) \\
 I_{l}^{{\rm SSP}} &=& -2.5 \log \left( \frac{\sum_{i} {\rm F}_{l,i}
{\rm N}_{i}} {\sum_{i} {\rm F}_{c,i} {\rm N}_{i}} \right)\,\,\,\ (Mag)
\label{int_ssp}
\end{eqnarray}

\noindent
where ${\rm N}_{i}$ is the number of stars in the bin.

When computing actual SSPs, single stars are identified to the
isochrone elemental bins defined in such a way that all relevant
quantities, i.e. luminosity, \logT, \logG\ and mass vary by small
amounts. In particular, the number of stars in an isochrone bin is
given by

\begin{equation}
{\rm N}_{i}=\int_{m_{a}}^{m_{b}}\phi(m)dm
\end{equation}

\noindent
where $m_{a}$ and $m_{b}$ are the minimum and maximum star mass in the
bin and $\phi(m)$ is the mass function in number. These are the
equations adopted to calculate the indices of SSPs.

Finally, in view of the discussion below it is worth reminding the
reader the definition of two indices that are commonly used but which
do not belong to the original Lick system. They are \MFe\ and \MgFe\

\begin{eqnarray}
\MFe     & = & 0.5\times (Fe5270+ Fe5335)  \nonumber
\end{eqnarray}

\begin{eqnarray}
\MgFe        & = & \sqrt{\mgb \times (0.5 \times Fe5270+0.5 \times Fe5335)} \nonumber
\end{eqnarray}

\subsection{Stellar models and isochrones}\label{mod_iso}

For the purposes of this study, we adopt the Padova Library of stellar
models and companion isochrones according to the release by
\citet{Girardi00}. This set of stellar models/isochrones differs
from the classical one by \citet{Bertelli94} for the efficiency of
convective overshooting and the prescription for the mass-loss rate
along the asymptotic red giant branch (AGB) phase. The stellar models
extend from the zero-age mean sequence (ZAMS) up to either the start
of the thermally pulsing AGB phase (TP-AGB) or carbon ignition. No
details on the stellar models are given here; they can be found in
\citet{Girardi00} and \citet{Girardi02}. Suffice it to mention that:
(i) in low-mass stars passing from the tip of red giant branch (T-RGB)
to the horizontal branch (HB) or clump, mass-loss by stellar winds is
included according to the \citet{Reimers75} rate with $\eta$=0.45;
(ii) the whole TP-AGB phase is included in the isochrones with ages
older than 0.1\,Gyr according to the algorithm of \citet{GirBer98} and
the mass-loss rate of \citet{Vassiliadis93}; (iii) four chemical
compositions are considered as listed in Table~\ref{enh-deg}.

\begin{table}
\small
\begin{center}
\caption[]{Chemical composition as a function of metallicity for the
SSPs in use.} \label{enh-deg}
\begin{tabular*}{36mm}{c c c }
\hline \multicolumn{1}{c}{Z} & \multicolumn{1}{c}{Y} &
\multicolumn{1}{c}{X}  \\
\hline
 0.008& 0.248 & 0.7440  \\
 0.019& 0.273 & 0.7080  \\
 0.040& 0.320 & 0.6400  \\
 0.070& 0.338 & 0.5430  \\
\hline
\end{tabular*}
\end{center}
\end{table}

\begin{figure}
\psfig{file=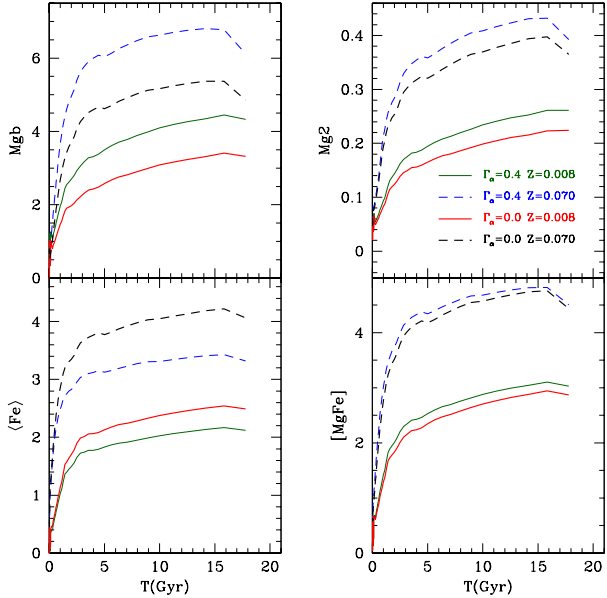,width=8.0truecm}
   \caption{Indices on the Lick resolution calculated with the
   $\mathcal{FF}$s and the new $\mathcal{RF}$s. The panels show the
   temporal variation of four different indices.}
\label{ssp_ff_nrf}
\end{figure}

\begin{figure}
\psfig{file=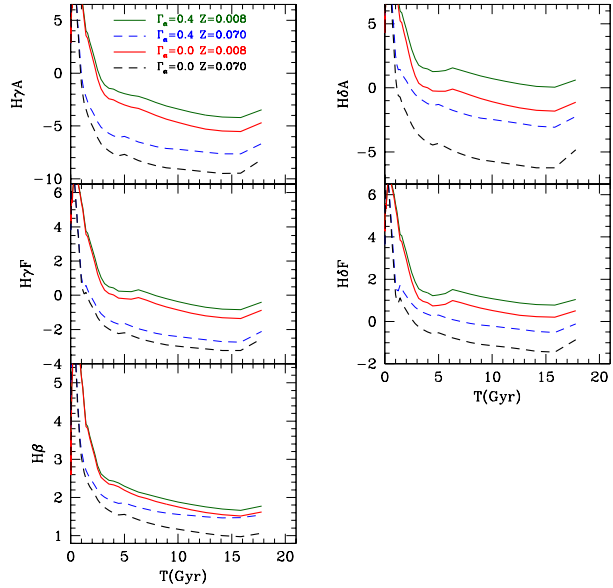,width=8.0truecm}
   \caption{The same as Fig.~\ref{ssp_ff_nrf} but for the Balmer Lick indices.}
\label{ssp_ff_nrf_B}
\end{figure}

\subsection{Results for SSP indices on the Lick resolution}\label{ind_ssp_ff}

Going into a detailed description of the dependence of the
absorption-line indices for SSPs on age, chemical composition and
\gaa, is beyond the scope of this study\footnote{The complete grids
of absorption-line indices are available from the authors upon request
and/or downloaded from the web site http://dipastro.pd.astro.it/galadriel.}. 
Suffice it to show in Figs.~\ref{ssp_ff_nrf} and \ref{ssp_ff_nrf_B}
the temporal evolution of nine important indices, i.e. \mgb, \mgii,
\MFe, \MgFe, and the Balmer Lick \Hbeta, \hgf, \hdf, \hga\ and \hda,
for the following combinations of metallicity and \gaa, namely Z=0.008
(solid lines) and Z=0.070 (broken lines), \gaa=0. (heavy lines) and
\gaa=0.4 (light lines). The age goes from 0.01 to 18\,Gyr. These
results are very similar albeit not identical to those found by TC04
adopting the same library of stellar spectra, the \citet{Worthey94}
$\mathcal{FF}$s, the old $\mathcal{RF}$s of \citet{Tripicco95} and the
algorithm of \citet{Trager20a}. Finally, we note that the most
controversial result in the studies by TC04 and \citet{Tantalo04b},
i.e. the increase of \Hbeta\ with \asfe\ (and/or $\Gamma_{Z}$ and/or
\gaa) at fixed age and metallicity, is recovered by the present
analysis.

\begin{figure}
\centerline{
\psfig{file=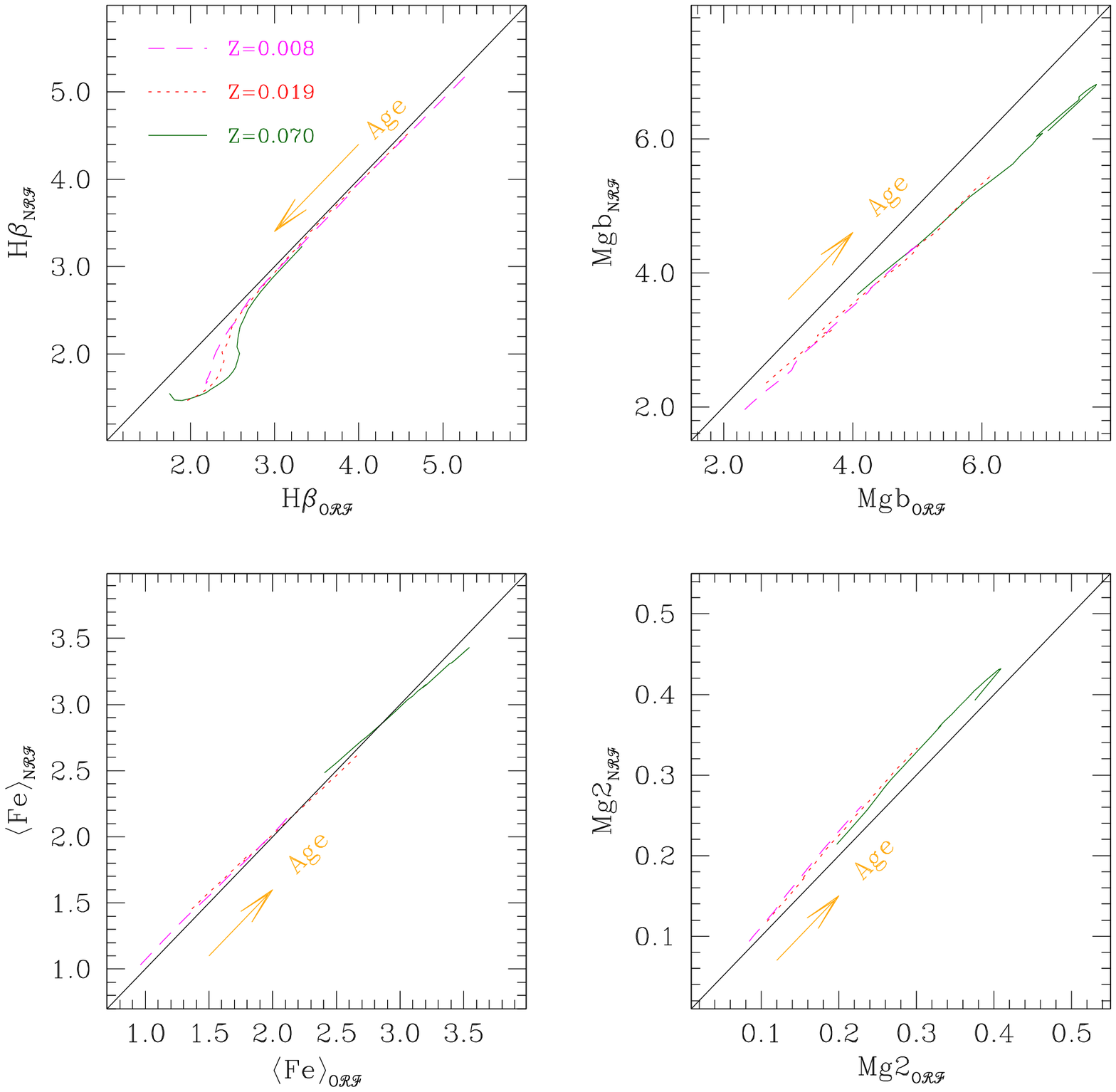,width=9.0truecm,height=9.0truecm}}
   \caption{Indices on the Lick resolution: comparison between \Hbeta\
   (top-left panel), \mgb\ (top-right panel), \MFe\ (bottom-left
   panel) and \mgii\ (bottom-right panel) calculated with the
   $\mathcal{FF}$s of \citet{Worthey94} but different $\mathcal{RF}$s
   and definition of enhancement. The indices shortly indicated as
   $I_{{\rm O}\mathcal{RF}}$ are calculated with the first definition
   of enhancement ($\Gamma_{Z}$=0.25) the old $\mathcal{RF}$s by
   \citet{Tripicco95} and the algorithm of equation~(\ref{dind}). The
   indices whereas $I_{{\rm N}\mathcal{RF}}$ are calculated with the
   second definition of enhancement (\gaa=0.4) the $\mathcal{RF}$s and
   equation~(\ref{ind_diff_enha}). Both groups have the same elements
   enhanced by \asfe=+0.4\,dex. The corresponding abundances are given
   in Tables~\ref{tab-enh} and \ref{tab-enh_zevar}. Finally, three
   different metallicities are considered as indicated.}
\label{comp_tb95_hrs}
\end{figure}

\subsection{Passing from old to new $\mathcal{RF}$s}\label{compar}

Comparing the present results with previous ones in literature is not
an easy task because the strict analogues do not exist. In brief, the
models by \citet{Trager20a,Trager20b}, \citet{Maraston03},
\citet{Thomas03}, \citet{Thomas03a}, \citet{ThoMara03} and TC04 are
calculated with the definition of enhancement at constant Z, the
classical $\mathcal{FF}$s of \citet{Worthey94}, the old
$\mathcal{RF}$s of \citet{Tripicco95} and either equation~(\ref{dind})
or (\ref{dind_thomas}) to pass from solar-scaled to $\alpha$-enhanced
mixtures. The present models adopt the definition of enhancement at
increasing Z, the classical $\mathcal{FF}$s of \citet{Worthey94} the
new $\mathcal{RF}$s calculated in this work and
equation~(\ref{ind_diff_enha}) to pass from solar-scaled to $\alpha$
enhanced indices.

In addition to this, there are differences arising from the quality of
the spectra for the calibrating stars adopted by \citet{Tripicco95}
and \citet{Munari05}. Owing to the many improvements in the physics of
stellar atmospheres introduced by \citet{Munari05} and
\citet{CastelliKurucz04}, the more recent compilations for the solar
abundances and many other details that are shortly summarized in
Section~\ref{high_res_sp}, the new $\mathcal{RF}$s likely supersede
the old ones.

Despite these major drawbacks, it might be useful to compare the
present models, shortly indicated as $I_{{\rm N}\mathcal{RF}}$, with
those calculated with the old procedure, shortly indicated as $I_{{\rm
O}\mathcal{RF}}$. These latter models have been explicitly calculated
with the abundances listed in Table~\ref{tab-enh}, i.e. by using the
same enhancement (\asfe=+0.4\,dex) but the definition at constant Z
(i.e. $\Gamma$=0.25 and \gaa=0.4), the \citet{Tripicco95}
$\mathcal{RF}$s and equation~(\ref{dind}).

For the sake of brevity, we compare four indices, namely \Hbeta,
\mgb, \MFe\ and \mgii, for three metallicities (Z=0.008, 0.019 and
0.070) and \gaa\ goes from 0 to 0.4 (the same values hold good for
both definitions of total enhancement). This choice for the
metallicity will allow us to check cases for which the $\mathcal{RF}$s
have been extrapolated. Furthermore passing from the $I_{{\rm
O}\mathcal{RF}}$ to the $I_{{\rm N}\mathcal{RF}}$ at given Z we have
also taken into account the correlation between \gaa\ and Z (and/or
[M/H]) discussed in Section~\ref{enha} (see Table~\ref{enh_chem}). The
correlations between $I_{{\rm O}\mathcal{RF}}$ and $I_{{\rm
N}\mathcal{RF}}$ are shown in Fig.~\ref{comp_tb95_hrs}. In each panel
three groups of models are displayed whose metallicity increases going
from bottom-left to top-right, whereas the age increases as indicated.
Along each curve the age goes from 1 to 18\,Gyr as in Figs.~\ref{ssp_ff_nrf} 
and \ref{ssp_ff_nrf_B}.

For \MFe\ there is no significant difference between the two groups;
\mgb$_{{\rm N}\mathcal{RF}}$ is systematically lower than
\mgb$_{{\rm O}\mathcal{RF}}$ by 0.2\AA\ to as much as 0.4\AA\ at increasing
age and/or metallicity; \mgii$_{{\rm N}\mathcal{RF}}$ tends to become
larger than \mgii$_{{\rm O}\mathcal{RF}}$ at increasing age by as much
as approximately 15 per cent. We suspect that for both indices the
offset is due to differences in the stellar spectra that are difficult
to pin down.

Finally, in the case of \Hbeta\ there is coincidence independently of
the metallicity and definition of enhancement for ages younger than
3\,Gyr, whereas for older ages \Hbeta$_{{\rm O}\mathcal{RF}}$ is
larger than \Hbeta$_{{\rm N}\mathcal{RF}}$ and the difference
increasing with metallicity. In the age range where the largest
deviation occurs, the maximum difference amounts to approximately 30
per cent. Careful inspection of the data reveals that the difference
mostly originates from correcting algorithm in usage when applied to
elements for which the $\mathcal{RF}$ is strong. Recalling that the
indices $I_{{\rm O}\mathcal{RF}}$ are calculated with [Ti/Fe]=0.40
(that corresponds to [Ti/H]=+0.1541), one of the elements with the
strongest $\mathcal{RF}$ in \citet{Tripicco95} tabulation, the excess
of \Hbeta$_{{\rm O}\mathcal{RF}}$ over \Hbeta$_{{\rm N}\mathcal{RF}}$
is most probably due to the correcting algorithm of
equation~(\ref{dind}) which tends to diverge whenever [$X_{i}$/Fe] or
[$X_{i}/{\rm H}$]/0.3 and ${\rm R}_{0.3}(X_{i})$ are high (see
equation~\ref{dind}). In the case of \Hbeta$_{{\rm N}\mathcal{RF}}$
the correction with respect to solar values is smaller (even though
sizable) because of the linearity of the algorithm in usage. For more
details on this subject the reader is referred to the discussion by
TC04.

From this comparison we learn that:

\noindent 
(i) In general, there seems to be overall agreement between the two
groups of indices despite the large difference in the definition of
enhancement. As the case at increasing Z requires the re-scaling of
the metallicity to lower values, the results eventually get close to
those at constant Z which implies a decrease in all elements that are
not enhanced (Fe, in particular).

\noindent
(ii) The new $\mathcal{RF}$s take into account effects that would
otherwise be missed, such as the dependence on the metallicity.

\noindent
(iii) The new $\mathcal{RF}$s for some indices and some stars, e.g.
\Hbeta\ for cool-dwarfs, have the opposite dependence on \gaa\ as
compared to that predicted by \citet{Tripicco95}. The result stems
from the modern physical input of the new spectra.

\noindent
No better comparison can be made at the present time, because we would
need stellar spectra with abundances of elemental species increased
one-by-one to derive the strict analogue of the $\mathcal{RF}$s by
\citet{Tripicco95} and  spectra with enhanced composition according to the
constant Z scheme. All this is still out of reach for obvious reasons.

\begin{figure}
\centerline{
\psfig{file=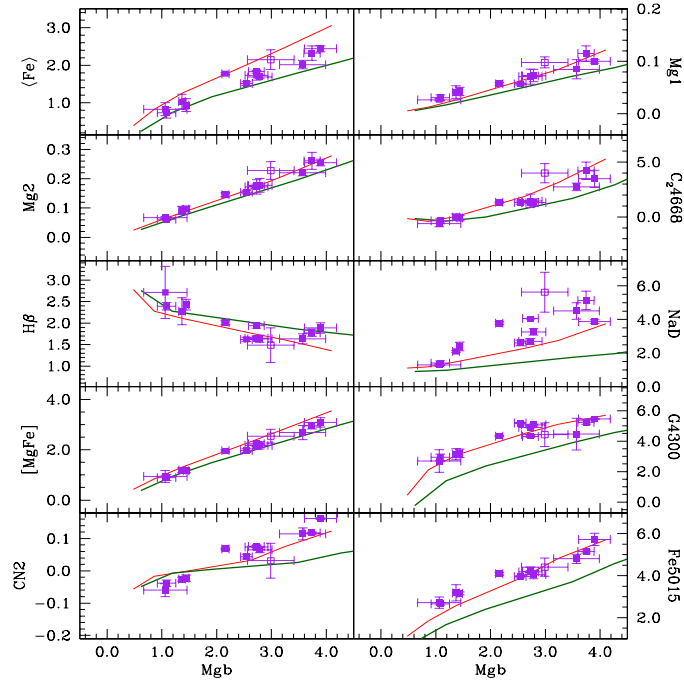,width=9.0truecm,height=9.0truecm}}
   \caption{The \mgb\ index versus ten different indices on the Lick
   resolution calculated with metallicities typical of globular
   clusters. The thin-solid lines are the models with solar-scaled
   pattern of abundances, whereas the thick-solid are the models with
   \gaa=0.4. Along each line are shown the indices of SSPs with the
   same age, $\sim$12\,Gyr and different metallicity from ${\rm
   [Z/Z_{\odot}]}$=--2 (Z$\sim$0.0001) to 0 (Z$\sim$0.019). These
   models are calculated with the classical $\mathcal{FF}$s and the
   new $\mathcal{RF}$s. The filled-squares are the globular cluster
   data, the open-square is the galactic bulge from \citet{Puzia02}.}
\label{globular_ff}
\end{figure}

\section{Comparison with globular clusters}\label{data_globular_cluster}

It is worth checking whether the present indices can reproduce the
data for the globular clusters.

To this aim we have taken the sample of galactic globular clusters and
bulge compiled by \citet{Puzia02} for which indices on the Lick system
have been measured. To analyze this sample we adopt the isochrones of
the Padova Library with chemical composition typical of the galactic
globular clusters \citep{Girardi00}, and derive the indices using the
classical $\mathcal{FF}$s and the new $\mathcal{RF}$s as described in
Section~\ref{ind_ssp}. The metallicity-enhancement relation is taken
into account when comparing indices for solar-scaled and
$\alpha$-enhanced mixtures.  The results are shown in
Fig.~\ref{globular_ff} for a number of indices. Each line show the
expected correlation between two indices at fixed age and varying
metallicity. The age under consideration is that typical of globular
clusters, i.e. 12 Gyr.  The metallicity increases from ${\rm
[Z/Z_{\odot}]}$=--2.0 (Z$\sim$0.0001) to ${\rm [Z/Z_{\odot}]}$=0
(Z$\sim$0.019) moving from bottom left to top right. The thin-solid
lines are the models with solar-scaled pattern of abundances, whereas
the thick-solid are the models with \gaa=0.40. Despite the fact that
these indices were not particularly designed to match the globular
clusters, the overall agreement is satisfactory. The new indices are
of the same quality as those in TC04 and \citet{Thomas03}. In general,
the agreement is better for the solar-scaled mixture even if values of
\gaa\ in the range from 0 to 0.4 cannot be excluded. The index \nad\
depart from the observational value both in \citet{Thomas03}, TC04 and
here.

\begin{table*}
\normalsize
\begin{center}
\caption[]{Ages, metallicities and \asfe\ for the sample of Galactic
Globular Clusters by \citet{Puzia02}. Columns (2) through (5) are
estimates of the parameters based on the colour-magnitude diagrams
or spectroscopic measurements; columns (6) through (8) are the
results derived from the absorption line indices.} \label{tab_gc}
\small
\begin{tabular*}{99mm}{c| r c r c| r c c}
\hline
\multicolumn{1}{c|}{NGC} &
\multicolumn{1}{c}{[Fe/H]$_{CG}$} &
\multicolumn{1}{c}{\asfe$^{d}$} &
\multicolumn{1}{c}{T$_{CG}$} &
\multicolumn{1}{c|}{$\sigma_{T}$} &
\multicolumn{1}{c}{[Z/H]} &
\multicolumn{1}{c}{\asfe} &
\multicolumn{1}{c}{Age} \\
\hline
5927 & --0.64$^{a}$ &  --- & 10.0 & 0.7$^{a}$ & --0.62 & 0.34 & 12.6 \\
6218 & --1.14$^{a}$ & 0.35 & 10.0 & 0.9$^{a}$ & --1.76 & 0.30 &  7.9 \\
6284 & --1.13$^{a}$ &  --- &  9.5 & 0.4$^{a}$ & --1.51 & 0.37 &  8.9 \\
6356 & --0.89$^{a}$ &  --- & 10.0 & 2.0$^{b}$ & --0.94 & 0.37 & 12.6 \\
6388 & --0.74$^{a}$ &  --- & 10.6 & 2.0$^{b}$ & --0.91 & 0.06 & 10.0 \\
6528 &   0.07$^{b}$ & 0.11 & 10.0 & 2.0$^{b}$ & --0.03 & 0.16 & 12.6 \\
6553 & --0.06$^{b}$ & 0.19 & 10.0 & 2.0$^{b}$ & --0.03 & 0.03 & 12.6 \\
6624 & --0.70$^{b}$ &  --- & 10.6 & 1.4$^{c}$ & --0.91 & 0.30 & 12.6 \\
6637 & --0.78$^{a}$ &  --- &  9.9 & 1.1$^{a}$ & --1.03 & 0.37 & 12.6 \\
6981 & --1.21$^{a}$ &  --- &  8.7 & 0.9$^{a}$ & --1.74 & 0.34 & 10.0 \\
\hline
\multicolumn{8}{l}{$^{a}$ \citet{DeAngeli05}} \\
\multicolumn{8}{l}{$^{b}$ \citet{Santos04}} \\
\multicolumn{8}{l}{$^{c}$ \citet{Salaris02}} \\
\multicolumn{8}{l}{$^{d}$ \citet{Pritzl05}} \\
\end{tabular*}
\end{center}
\end{table*}

To further confirm the quality of the theoretical indices we derive
the ages, metallicities and enhancement degree from the indices for
the sub-set of this sample of globular cluster for which the same
parameters are known from the colour-magnitude diagrams.

To this aim, first we select the globular clusters for which estimate
of ages and metallicities (i.e. [Fe/H]) have been derived from the
colour-magnitude diagrams
\citep[see][]{DeAngeli05,Santos04,Salaris02,Schiavon05}. Then we
consider the current estimates of enhancement (i.e. \asfe) obtained
directly from stars and take these as indicators of the total degree
of enhancement in each cluster \citep[see]{Pritzl05}.
Table~\ref{tab_gc} lists the metal content, degree of enhancement and
age together with its uncertainty we have been able to collect from
literature (the sources of data are indicated). These are the values
we would like to recover from the indices.

To derive metallicities, degree of enhancement and age from indices we
adopt the {\it Minimum-Distance Method} originally proposed by
\citet[][]{Trager20a} and amply described in TC04. Due to the
different sensitivity of the adopted indices to each parameter we
prefer to use the so-called {\it Recursive Minimum-Distance Method}
\footnote{The method was introduced and applied in the first version
of this paper by \citet{Tantalo04a} appeared on {\it
http://xxx.lanl.gov/abs/astro-ph/0305247v1} to which the reader should
refer for details.}.

Taking advantage of the high sensitivity of \MFe\ and \mgii\ to \asfe\
and their lower sensitivity to age and metallicity, we have apply the
minimum-distance method to derive \asfe\ form the plane \MFe-\mgii\
assuming the provisional age T$\sim$13. This value, shortly indicated
as $\Gamma^{0}_{\alpha}$, is used to construct the functions
\Hbeta(T,Z,$\Gamma^{0}_{\alpha}$), \MFe(T,Z,$\Gamma^{0}_{\alpha}$) and
\mgb(T,Z,$\Gamma^{0}_{\alpha}$) and to solve the equations:

\begin{displaymath}
\Hbeta_{obs}=\Hbeta(T,Z,\Gamma^{0}_{\alpha})
\end{displaymath}
\begin{displaymath}
\MFe_{obs}=\MFe(T,Z,\Gamma^{0}_{\alpha})
\end{displaymath}
\begin{displaymath}
\mgb_{obs}=\mgb(T,Z,\Gamma^{0}_{\alpha})
\end{displaymath}

\noindent
for the age and metallicity (T and Z). The procedure is iterated till
full consistency is achieved.

The results of this analysis are shown in Table~\ref{tab_gc} (columns
(6) through (8) for the metallicity, \asfe\ and age,
respectively). Metallicities, degree of enhancement and ages derived
from indices fully agree with those obtained from the colour-magnitude
diagrams. This holds good both for individual clusters and the mean
values for the whole sample: $\langle$Age$\rangle$=11.2$\pm$1.8,
$\langle$[Z/H]$\rangle$=--0.95$\pm$0.62\footnote{We pass from [Z/H] to
[Fe/H] with the aid of the obvious relation:
[Fe/H]=[Z/H]--$\Gamma_{Z}$, which with the definition of enhancement
adopted in this study is [Fe/H]=[Z/H]--0.625\gaa.} and
$\langle$\asfe$\rangle$=0.26$\pm$0.13 \footnote{We can directly
compare our results with the observed ones due to the fact that these
latter are generally the average of the [Mg/Fe], [Si/Fe], [Ca/Fe] and
[Ti/Fe] abundance ratios.}.

\section{Discussion and conclusions}\label{conclu}

We have generated synthetic absorption-line indices on the Lick system
based on the recent library of 1-\AA\ resolution spectra calculated by
\citet{Munari05} over a large range of atmospheric parameters
(\logT, \logG\ and [Fe/H]) and both for solar and $\alpha$-enhanced
abundance ratios in the chemical composition. The main results of this
study are:\\

\noindent
(i) First we derive a modern version of the so-called $\mathcal{RF}$s
of \citet{Tripicco95}. Contrary to the previous situation in which the
$\mathcal{RF}$s were known only for three stars of given \logT\ and
\logG, now the $\mathcal{RF}$s are given for large range ranges of \logT,
\logG\ and [Fe/H] (or ${\rm [Z/Z_{\odot}]}$). Not only the $\mathcal{RF}$s
vary with the type of star but also with the metallicity\footnote{The
new $\mathcal{RF}$s are given as large 3D-matrices made available on
the web page http://dipastro.pd.astro.it/galadriel.}. The effect of
metallicity is important and cannot be neglected. While completing
this study, similar analysis has been published by \citet{Korn2005}
who strictly following the \citet{Tripicco95} approach, presented the
$\mathcal{RF}$s for three stars (a dwarf, a turn-off and a red giant
for six chemical composition). Their results are similar to ours even
if the coverage of the effective temperature-gravity intervals for the
reference stars is much narrower.\\

\noindent
(ii) With the aid of the new $\mathcal{RF}$s and the $\mathcal{FF}$s
of \citet{Worthey94} indices for SSPs are calculated and compared with
the old ones by TC04. But for the differences caused by the new
$\mathcal{RF}$s and the kind of enhancement adopted by TC04 and here,
the agreement is good thus confirming that the method adopted by TC04
to account for the effect of $\alpha$-enhancement albeit hampered by
the coarse grid of calibrating stars in \citet{Tripicco95} was
correct. The present results clearly demonstrate that not only all
indices depend on the enhancement but also that \Hbeta\ increases with
it as already anticipated by TC04 and contrary to what claimed by
\citet{Thomas03,Thomas03a} and only marginally admitted by \citet{Korn2005}
in their recent study.\\

\noindent
(iii) The new indices have been compared to those of galactic globular
clusters for which there are independent estimates of age, metallicity
and degree of enhancement and used to re-derive these basic parameters
by means of the recursive minimum distance method customarily applied
to galaxies. The estimates of the three key parameters from the two
methods agree each other thus suggesting that the indices for SSPs we
have calculated are correct.

\begin{acknowledgements}
This study has been financed by the Italian Ministry of Education,
University, and Research (MIUR), and the University of Padua under the
special contract `Formation and evolution of elliptical galaxies: the
age problem'.
\end{acknowledgements}

\bibliographystyle{apj}          
\bibliography{mnemonic,biblio}    

\appendix

\section{}
Tables~A.1--A.5 list the new $\mathcal{RF}$s calculated with the
1-\AA\ resolution spectra from \citet{Munari05} as function of \Teff\
and \logG\ and for four different metallicities: ${\rm
[Z/Z_{\odot}]}$=--2.0, --1.5, --1.0, --0.5. For each index we give the
value $I_{{\rm sol}}$ for the solar abundance ratios and the
difference $\left(\delta I\right)$=$I_{{\rm enh}} - I_{{\rm sol}}$
between the $\alpha$-enhanced and the solar case. The corresponding
$\mathcal{RF}$s can immediately be derived from relation (\ref{r04}).
Tables~A.6--A.10 are the same but for the high-resolution spectra
degraded to the Lick resolution.

\end{document}